\documentclass{article}
\usepackage[a4paper, total={6in, 8in},margin=2.5cm]{geometry}
\usepackage{amsmath,amssymb,amsthm, amsfonts,latexsym}
\usepackage[T1]{fontenc}
\usepackage[english]{babel}
\usepackage[utf8]{inputenc}
\usepackage{mathtools}
\usepackage[breaklinks,hidelinks]{hyperref}
\usepackage{cite}
\usepackage{cleveref}
\usepackage{authblk}
\usepackage{comment}
\usepackage{lineno}
\usepackage{setspace}
\usepackage{xcolor}
\usepackage{subcaption}
\usepackage{tikz}
\usetikzlibrary{arrows.meta,positioning,calc,shapes.geometric,fit}

\usepackage{algorithm}
\usepackage{algpseudocode}

\usepackage{float,afterpage}
\usepackage{blkarray}

\usepackage{environ}

\usepackage{graphicx}

\usepackage{booktabs}
\usepackage[scale=2]{ccicons}
\usepackage{amsmath,amsfonts}
\usepackage{xspace}

\title{Flexible inference of evolutionary accumulation dynamics using uncertain observational data}

\author[1,*]{Jessica Renz}
\author[1]{Morten Brun}
\author[1,2]{Iain G. Johnston}

\date{ }
\affil[1]{\small Department of Mathematics, University of Bergen, Bergen, Norway}
\affil[2]{\small Computational Biology Unit, University of Bergen, Bergen, Norway}
\affil[*]{\small correspondence to \url{jessica.renz@mis.mpg.de}}

\begin{document}

\maketitle

\begin{abstract}
Understanding and predicting evolutionary accumulation pathways is a key objective in many fields of research, ranging from classical evolutionary biology to diverse applications in medicine. In this context, we are often confronted with the problem that data is sparse and uncertain, which challenges many existing approaches for inferring evolutionary pathways. To use the available data as powerfully as possible, inference approaches that can handle this uncertainty are required. In this article we introduce HyperLAU, a new algorithm for learning evolutionary pathways for accumulation processes from datasets with substantial uncertainty. Expanding the flexibility of accumulation modelling, HyperLAU allows us to infer dynamic pathways and interactions between features, even when large sets of particular features are unobserved across the source dataset. We show that HyperLAU is able to identify evolutionary pathways found by other tools that rely on completely specified data, even when up to 50\% of the features in the input data are artifically obscured. Additionally, we demonstrate how it can help to overcome possible biases that can occur when reducing the used data by excluding uncertain subsets. We illustrate the approach with case studies on the evolution of multidrug resistance in tuberculosis, $C_4$ photosynthesis in plants, and mitochondrial reduction across eukaryotes, showing in each case that HyperLAU allows more flexible data, strengthens scientific interpretation, and provides new information about evolutionary pathways compared to existing methods.
\end{abstract}

\section{Introduction}
Many processes in the natural sciences and especially in biology and medicine can be modelled as the accumulation or loss of binary features \cite{diaz-uriarte_picture_2023, schill_reconstructing_2024, diaz-uriarte_evam_2022}. Common examples include evolutionary processes, such as accumulation of mutations in cancer \cite{ schwartz_evolution_2017, diaz-uriarte_every_2019, beerenwinkel_cancer_2015, takeshima_accumulation_2019, moen_hyperhmm_2023}, gene loss in mitochondria \cite{johnston_evolutionary_2016, maier_massively_2013}, mutations in bacteria \cite{nichol_steering_2015, tan_hidden_2011, greenbury_hypertraps_2020, aga2025natural}, and the appearance of phenotypic characters \cite{ moen_hyperhmm_2023, nichol_steering_2015}. Other examples of accumulation processes include, for example, patients acquiring symptoms in progressing diseases \cite{ johnston_precision_2019, schill_overcoming_2024}. Understanding and predicting such processes can contribute both to the basic knowledge of the mechanisms involved, and to potential application -- for example, in triaging patients \cite{ johnston_precision_2019} and determining optimal treatments \cite{aga_hypertraps-ct_2024, renz_evolutionary_2024}.\\

 Such approaches include models that assume a tree structure like  Oncogenetic Trees (OT) \cite{desper_inferring_1999, szabo_estimating_2002}, and the Mutation Order (MO) model \cite{gao_phylogenetic_2022}. A generalization giving up the tree structure is given by conjunctive Bayesian networks (CBN) \cite{beerenwinkel_evolution_2006, beerenwinkel_conjunctive_2007,gerstung_quantifying_2009,beerenwinkel_markov_2009,montazeri_large-scale_2016}. This approach is complemented by the disjunctive Bayesian networks \cite{nicol_oncogenetic_2021} where only one parent needs to be present that a mutation can happen, and DAG models where parental nodes are not a necessary requirement, but have a positive impact to direct descendant nodes \cite{hjelm_new_2006}. The most recent class of models considers dependencies between states rather than single events, allowing the occurrence of features in all possible orderings. This group includes HyperTraPS \cite{johnston_evolutionary_2016, greenbury_hypertraps_2020,aga_hypertraps-ct_2024}, Mutual Hazard Networks (MHN) \cite{schill_modelling_2020, rupp_differentiated_2024, schill_overcoming_2024,gotovos_scaling_2021,chen_timed_2023, luo_joint_2023}, HyperHMM \cite{moen_hyperhmm_2023}, and HyperMk \cite{johnston_hypercubic_2024}. Some overviews over the recent methods for modelling accumulation and loss of features are given in \cite{diaz-uriarte_picture_2023, renz_evolutionary_2024}. \\
	
	Several papers have already addressed the problem of estimating a transition matrix of a Markov chain, when observations for some steps in the chain are missing \cite{barsottiEstimatingTransitionMatrix2014,demboParameterEstimationPartially1986,rabbatNetworkInferenceCoOccurrences2008,barrattFittingFeaturedependentMarkov2023,yehEstimatingTransitionProbabilities2010,leskelaRobustEstimationMarkov2026,sherlaw-johnsonEstimatingMarkovTransition1995}. Also, many of the named models from above are concerned with exactly this task. However, in several scientific and medical contexts, observational data on evolutionary accumulation processes often additionally include uncertainties and unknown information regarding individual features. This corresponds to structured uncertainty over which subset of states the system occupies at a given time, combined with structured absent observations through chains of a fixed length running in the direction of monotonic accumulation. 
	 The extent to which existing models can handle this uncertainty and exploit these specific structures is limited. In most of the named models, there are only general correction mechanisms or error rates included, which can account for some noise. MO \cite{gao_phylogenetic_2022} explicitly allows missing or ambiguous information for specific features, but is subject to the infinite-sites assumption (every event occurs at most once, prohibiting parallel evolution), which is not appropriate in several scientific contexts. HyperTraPS-CT \cite{aga_hypertraps-ct_2024} does not have this assumption, but represents data as a set of ancestor-descendant states and can only deal with uncertainties in the descendant states, not in the ancestral ones. So far the only model that can fulfill all these requirements is the Mk-model \cite{johnston_hypercubic_2024} which, however, is limited to very few features due to the high computational runtime.\\

      Here, we introduce HyperLAU, an approach for accumulation modelling that has no underlying infinite site assumption but can deal with longitudinal input data, including specified uncertainties in ancestor and descendant states (see Figure \ref{graph_abstr}). On the same time, it can handle input data with more features than the Mk-model in a smaller amount of time. Borrowing the idea for the option to model different orders of dependencies from \cite{greenbury_hypertraps_2020}, we additionally keep this flexibility, constituting a further advantage of our model. In addition to case studies on diverse evolutionary topics from photosynthesis to organelle evolution, we will particularly motivate the approach in the context of anti-microbial resistance (AMR) evolution. AMR is one of the top threats to human health worldwide, with 4.95 million deaths associated with AMR in 2019 \cite{murray_global_2022}. Data on bacterial drug resistance is often either inferred from genome information (leading to uncertainty) or measured using laborious experiments (often leading to incompleteness). The ability to handle incomplete, phylogenetically-related data is therefore a substantial advantage to accumulation models in this context \cite{renz_evolutionary_2024, aga2025natural}.

\begin{figure}
	\centering 
	\begin{tikzpicture}[ 
		font=\small, 
		>=Latex, 
		panel/.style={ draw, rounded corners, minimum width=3.5cm, minimum height=5cm, align=center }, arrow/.style={ thick, ->} 
		]
		  \node[panel](A) at (0,0){

		  \begin{tabular}{ccccccc} 
		  0 & ? & 1 & & 1 & ? & 1\\ 
		  ? & 1 & 1 & & 1 & 1 & 1\\ 
		  0 & 0 & ? & & 0 & 0 & ?
		  \end{tabular} \\[0.4cm]

		  Ancestor--descendant\\ 
		  pairs with missing\\ 
		  observations };
		  	\node[font=\bfseries] at (A.north) [yshift=-0.45cm] {A. Input data};
		   \node[panel,right=1.2cm of A, minimum width=4.5cm] (B) {}; 
		   
		   \node[font=\bfseries] at (B.north) [yshift=-0.45cm] {B. Compatible states}; 
		  \begin{scope}[shift={(5.2,0.3)}] 
		  	
		  	\coordinate (v1) at (-0.6,-0.6); 
		  	\coordinate (v2) at (0.6,-0.6); 
		  	\coordinate (v3) at (0.6,0.6); 
		  	\coordinate (v4) at (-0.6,0.6); 
		  	\coordinate (v5) at (-0.2,-0.2); 
		  	\coordinate (v6) at (1.0,-0.2); 
		  	\coordinate (v7) at (1.0,1.0); 
		  	\coordinate (v8) at (-0.2,1.0); 
		  	
		  	\draw[gray] (v1)--(v2)--(v3)--(v4)--cycle; 
		  	\draw[gray] (v5)--(v6)--(v7)--(v8)--cycle; 
		  	\draw[gray] (v1)--(v5); 
		  	\draw[gray] (v2)--(v6); 
		  	\draw[gray] (v3)--(v7); 
		  	\draw[gray] (v4)--(v8); 
		  	
		  	\node[circle,fill=gray!40,inner sep=2pt] at (v1) {}; 
		  	\node[circle,fill=blue!70,inner sep=2pt] at (v4) {}; 
		  	\node[circle,fill=gray!40,inner sep=2pt] at (v6) {}; 
		  	\node[circle,fill=red!70,inner sep=2pt] at (v7) {}; 
		  	\node[circle,fill=gray!40,inner sep=2pt] at (v2) {}; 
		  	\node[circle,fill=red!70,inner sep=2pt] at (v3) {}; 
		  	\node[circle,fill=gray!40,inner sep=2pt] at (v5) {}; 
		  	\node[circle,fill=blue!70,inner sep=2pt] at (v8) {}; 
		  	
		  	\end{scope} 
		  	
		  	\node[align=center] at (B.center) [yshift=-1.1cm] 
		  	{ Construct vectors of states\\ 
		  	compatible with observations }; 
		  	
		  	\node[align=center] at ($(B.south)+(0,0.6)$) 
		  	{ 
		  	{\color{blue}$c^a=(0,1,0,1,0,0,0,0)$}\\ 
		  	{\color{red}$c^d=(0,0,0,0,0,1,0,1)$} 
		  	}; 
		  	
		  	
		  	\node[panel,right=1.2cm of B, minimum width=4.7cm] (C) {}; 
		  	
		  	\node[font=\bfseries] at (C.north) [yshift=-0.45cm] 
		  	{C. Likelihood evaluation}; 
		  	
		  \node[ draw, rounded corners, minimum width=0.9cm, minimum height=0.5cm ] (A1) at ($(C.center)+(-1.3,1.4)$) {$A$}; 
		  \node[ draw, rounded corners, minimum width=0.9cm, minimum height=0.5cm ] (ca) at ($(C.center)+(-1.3,0.5)$) {$c^a$}; 
		  \node[ draw, rounded corners, minimum width=0.9cm, minimum height=0.5cm ] (A2) at ($(C.center)+(-1.3,-0.4)$) {$A$}; 
		  \node[ draw, rounded corners, minimum width=0.9cm, minimum height=0.5cm ] (cd) at ($(C.center)+(-1.3,-1.3)$) {$c^d$}; 
		  \node[ draw, rounded corners, fill=blue!10, minimum width=1.8cm, minimum height=0.5cm ] (L) at ($(C.center)+(-1.3,-2.2)$) {Likelihood}; 
		  
		  \draw[->,thick] (A1) -- (ca); 
		  \draw[->,thick] (ca) -- (A2); 
		  \draw[->,thick] (A2) -- (cd); 
		  \draw[->,thick] (cd) -- (L); 
		  \node[align=center,font=\scriptsize] at ($(ca.east)+(1.4,0)$) {ancestor\\constraint}; 
		  \node[align=center,font=\scriptsize] at ($(A1.east)+(1.4,0)$) {state\\propagation}; 
		  \node[align=center,font=\scriptsize] at ($(A2.east)+(1.4,0)$) {state\\propagation}; 
		  \node[align=center,font=\scriptsize] at ($(cd.east)+(1.4,0)$) {descendant\\constraint};
		  	
		  	
		  	\coordinate (ABmid) at ($(A.center)!0.5!(B.center)$);
		  \node[panel] (D) at ($(ABmid)+(0,-6.5)$) {};
		  	
		  	\node[font=\bfseries] at (D.north) [yshift=-0.45cm] {D. Optimisation}; 
		  	
		  	\coordinate (o) at ($(D.center)+(0,0.5)$); 
		  	
		  	\draw[thick,->] 
		  	(o) +(130:0.9) 
		  	arc (130:20:0.9);
		  	
		  	\draw[thick,->] 
		  	(o) +(10:0.9) 
		  	arc (10:-100:0.9); 
		  	
		  	\draw[thick,->] 
		  	(o) +(-110:0.9) 
		  	arc (-110:-220:0.9); 
		  	
		  	\node at ($(D.center)+(0,0.5)$) 
		  	{SA}; 
		  	
		  	\node[align=center] at ($(D.center)+(0,-1.0)$) 
		  	{ 
		  	Simulated annealing\\ 
		  	maximises likelihood 
		  	}; 
		  	
		  	\node at ($(D.south)+(0,0.9)$) 
		  	{$L_0 \rightarrow L_1 \rightarrow ... \rightarrow L_{\max}$}; 
		  	
		  	
		  	\node[panel,right=1.2cm of D,minimum width=4.5cm] (E) {}; 
		  	\node[font=\bfseries] at (E.north) [yshift=-0.45cm] {E. Pathway summaries};
		  	
		  	\begin{scope}[shift={($(E.center)+(0.3,0)$)}] 
		   \node[circle,fill=gray!35,inner sep=2.2pt] (s000) at (-1.5,-1.0) {}; 
		   \node[circle,fill=gray!35,inner sep=2.2pt] (s100) at ( 0.0,-1.0) {}; 
		   \node[circle,fill=gray!35,inner sep=2.2pt] (s010) at (-1.5, 0.5) {}; 
		   \node[circle,fill=gray!35,inner sep=2.2pt] (s110) at ( 0.0, 0.5) {}; 
		   \node[circle,fill=gray!35,inner sep=2.2pt] (s001) at (-0.8,-0.3) {}; 
		   \node[circle,fill=gray!35,inner sep=2.2pt] (s101) at ( 0.7,-0.3) {}; 
		   \node[circle,fill=gray!35,inner sep=2.2pt] (s011) at (-0.8, 1.2) {}; 
		   \node[circle,fill=gray!35,inner sep=2.2pt] (s111) at ( 0.7, 1.2) {}; 
		   
		    \draw[gray!50] (s000)--(s100); 
		    \draw[gray!50] (s000)--(s010); 
		    \draw[gray!50] (s000)--(s001); 
		    \draw[gray!50] (s100)--(s110); 
		    \draw[gray!50] (s100)--(s101); 
		    \draw[gray!50] (s010)--(s110); 
		    \draw[gray!50] (s010)--(s011); 
		    \draw[gray!50] (s001)--(s101); 
		    \draw[gray!50] (s001)--(s011); 
		    \draw[gray!50] (s101)--(s111); 
		    \draw[gray!50] (s110)--(s111); 
		    \draw[gray!50] (s011)--(s111); 
		    
		    \draw[->,ultra thick,blue!75] (s000) -- (s100); 
		    \draw[->,ultra thick,blue!75] (s100) -- (s101); 
		    \draw[->,ultra thick,blue!75] (s101) -- (s111); 
		    
		     \draw[->,line width=1.5pt,blue!40] (s000) -- (s010); 
		     \draw[->,line width=1.5pt,blue!40] (s010) -- (s110); 
		     \draw[->,line width=1.5pt,blue!40] (s110) -- (s111); 
		     \end{scope} 
		     
		     \node[align=center,font=\small] at ($(E.south)+(0,0.7)$) {Dominant pathways\\and transition fluxes};
		  	 	
		  	 	\draw[arrow] (A.east) -- (B.west); 
		  	 	\draw[arrow] (B.east) -- (C.west); 
		  	 	\draw[arrow] (C.south) -- (D.north); 
		  	 	\draw[arrow] (D.east) -- (E.west); 
		  	 	\end{tikzpicture}

    \caption{\small \textbf{HyperLAU workflow.} Learning evolutionary trajectories on a hypercube, based on data that contains uncertainties. \textbf{(A)} Dataset (structure can be cross-sectional or longitudinal) containing information about the presence (1) or absence (0) of certain features represented as binary strings. A '?' indicates that no information for this features is available. \textbf{(B)} Construction of vectors that contain information about compatible states for the ancestor string and the descendant string of a data pair. A '1' indicates that the string in the dataset is compatible to the corresponding node, whereas a '0' indicates non-compatibility. \textbf{(C)} The probabilities obtained by matrix multiplications are evaluated first at the compatibility vector of the ancestor string, and later at the compatibility vector of the descendant string. With the help of this information, a likelihood function can be calculated. \textbf{(D)} A simulated annealing algorithm iteratively searches for transition matrices that optimise the likelihood function. \textbf{(E)} The inferred transition parameters provide information about dominant evolutionary pathways and transition fluxes.}  
    \label{graph_abstr}
\end{figure}

\section{Materials and methods}
\textbf{HyperLAU.} We consider a state space $S$ that is the set of all binary strings of length $L$, where $L$ is the number of features in our system. Each binary string describes a state by the presence (1) or absence (0) of each of these $L$ features. A transition from state $s_i$ to $s_j$ is allowed if the binary string $s_j$ differs from $s_i$ by the presence of exactly one `1': that is, only transitions that involve the acquisition of one feature are allowed. The resulting structure is a directed acyclic graph which we call a hypercubic transition graph. The HyperLAU algorithm can be used to learn the transition probabilities $a_{s_i,s_j} = P(X_{t} = s_i | X_{t-1} = s_j) $ of a discrete Markov process $X_t$ on such a hypercubic transition graph. In so doing, the likely evolutionary pathways supported by observation are inferred.\\

We consider observations of the general form $\{k^{a}, k^{d}\}$, which are ancestor ($a$) and descendant ($d$) pairs derived from a dataset (see below). Each member of a pair is a string of length $L$ which consists of known (0 or 1), or uncertain (?) markers for each feature. A state node $s$ in the hypercube is called \emph{compatible} with an observation $k$ if and only if all positions in the string that are not represented by a uncertainty marker are equal. Our goal is a likelihood calculation finding the probability that a random walk that runs from the $000...$ state to the $111...$ state and emits exactly two signals will produce signals compatible with the ancestral and descendant observations in a datapoint. Beginning with an initial guess of a transition matrix, we then optimize the likelihood.\\

HyperLAU allows for inference of the transition parameters under different models for interaction or independence between features involved. The user can choose what level of interactions or influences between the different features they want to allow. This is done by an input parameter \texttt{model}$\in \{F,1,2,3,4\}.$ For \texttt{model} $=F$ (full), arbitrary interactions between all features are allowed. In this case, we learn a transition parameter for every allowed state-to-state transition by itself. We consider the whole transition matrix directly as the object we change in order to optimize the likelihood. For the four other models, state-state transition probabilities are functions of the changing feature and a set of contributions from acquired features, as also used in \cite{aga_hypertraps-ct_2024}. This means our parameter set consists of a base rate with which every feature occurs, and for \texttt{model}$\in \{2,3,4\}$ additional parameters that reflect the influence that the existence of a set of features can have on the rate with which another feature is obtained. In \texttt{model} = $1$, there is only the base rate and it does not matter what other features are already obtained. The other three models allow accordingly interactions of order two, three or four. In these models, influences between either single respectively pairs or triplets of already acquired features can increase or decrease the probability for other features to occur.  During the optimization procedure we make changes only on this parameter set. The transition matrix, which still is used in the calculation of the likelihood, can easily be created from this parameter set. \\

The core piece of the new method that is presented here, is the calculation of the log-likelihood of seeing the input data (including uncertainty markers) for a given transition or rate matrix (see Algorithm \ref{likelhood_code} in the SI). To take the uncertainty markers that can occur in the dataset into account, we have to consider all nodes in the hypercube that are compatible with the observations. To this end, for every observation $\{k^{a}, k^{d}\}$ in our dataset, the information of compatibility is stored in a vector $c^{a}$ for the ancestor and a vector $c^{d}$ for the descendant, where we set $c^{i}_{s} = 1$ if node $s$ is compatible with $k^{i}$ and 0 otherwise.  \\

Then the likelihood function of seeing a data point of $L$ features given a transition matrix $A$ can be described by the following formula (derived in the SI):
\begin{equation}
	\label{lik_one_point}
	\mathcal{L}((k^{a},k^{d})|A) =\frac{2}{(L+1)(L+2)}\cdot \sum_{t=0}^{L}\sum_{t'=t}^{L} c^{d}\cdot Q_{.,t'}^{t,a}
\end{equation}
with
\begin{equation*}
	Q_{.,t'}^{t,a} = A^{t'-t}\cdot \left(c^{a}\star P_{.,t}\right).
\end{equation*}
Here $P_{.,t}$ describes the probability distribution of occupancy for the different nodes of the hypercube at step $t$, and $\star$ is the entry-wise product of two vectors.\\

The factor $2/(L+1)(L+2)$ is the normalization constant reflecting the $(L+1)(L+2)/2$ possible combinations of nodes on the hypercube at which ancestor and descendant can be sampled. The detailed derivation and the implemented algorithm are described in the SI.\\

Calculating Equation (\ref{lik_one_point}) for every entry in our dataset, taking the logarithm, and summing up the obtained log-likelihoods, we get an overall logarithmic likelihood of obtaining the data in the dataset, given the transition matrix. This is also the function that needs to be optimized by adapting the transition or rate matrix. The final output is then the transition matrix that maximises the likelihood for our input data, providing the most likely transition probabilities for evolutionary trajectories on the hypercube.\\

We note that this calculation considers the likelihood contribution of all compatible pathways equally weighted. In cases where a feature's observations are dominated by uncertainty markers, this may lead to an inaccurate estimation of the prevalence of a feature in the dataset (see Discussion).\\

\textbf{Data to (independent) observations.} Considering ancestor-descendant pairs $\{k^{a}, k^{d}\}$ makes it possible to describe an evolutionary behaviour of a sample, either measured at two different time points (longitudinal data) or if they are genuinely phylogenetically-linked ancestor and descendant. Even cross-sectional data can be used, by setting all positions of all ancestor strings to zero. This reflects the general assumption that every modelled trajectory originally started at the zero-state where no features are obtained.\\

Consider the case where we have phylogenetically-embedded observations $0^L \rightarrow a, \ a\rightarrow b$ and $a\rightarrow c$, corresponding to observations of $0^L$ at the root of the tree, state $a$ at ancestral node $A$, and states $b$ and $c$ at two daughter tips $B$ and $C$. Treating emission probabilities as constants, we can describe the probability of obtaining these observations by:
\begin{equation*}
	P(\text{observe }0^{L}\rightarrow a,a\rightarrow b, a\rightarrow c)P(0^L\rightarrow a)P(a\rightarrow b)P(b\rightarrow c)
\end{equation*}
and as shown in \cite{johnston_evolutionary_2016} we can decompose all observations from a phylogeny into a similar product of independent transitions (one for each edge on the phylogeny). But if the state at ancestor $A$ is incompletely known we cannot decompose in this way. Instead we need to consider all possible latent states $a'$ of the system at state $A$:
\begin{equation*}
	P(\text{observe }0^L \rightarrow a, a\rightarrow b, a\rightarrow c)\sum_{a'}P(0^L\rightarrow a')P(a'\rightarrow b)P(a'\rightarrow c) 
\end{equation*}
which can no longer be written as a product of independent transition probabilities.\\

To avoid this issue, we prune transitions from our reconstructed phylogeny, first to consider only transitions to tip states, and then so that no pair of transitions in our final observation dataset have a most recent common ancestor that is inferred not to be $0^L$. In the example above, we would retain both $a\rightarrow b$ and $a\rightarrow c$ as elements of our observation set only if the inferred state at $A$ was compatible with $0^L$. We do this simply by pairwise comparison of the full set of transitions, and removing one transition from any pair where the MRCA of the two ancestor nodes has an inferred state incompatible with $0^L$.\\

To establish that our model is able to predict the most likely evolutionary pathways even with uncertain states in the data, we artificially introduced uncertainty into some datasets through random sample (see \ref{including_uncertainty}).\\

\textbf{Optimization and convergence.}  The optimization process follows a simulated annealing process. However, the core algorithm for the calculation of the likelihood can also be embedded in other optimisation methods. For detailed information about parameters and convergence see \ref{optimisation_convergence}. The runtime scales exponentially in the number of features. Also the proportion of uncertainty markers has an effect on the runtime, although less extreme. For a more detailed runtime analysis and a comparison of the results with other models, see \ref{subsec_analysis_comparison}. \\

\textbf{Bootstrapping.} For determining the uncertainty of our inference results we implemented a bootstrap option on the input data. The number of bootstrap resamples can be specified via an input parameter in the command line. All reported CVs in this article are based on 10 bootstrap resamples.

\section{Results}

\subsection{HyperLAU on toy examples}
\label{subsection_toy}
First we demonstrate the effect of our algorithm on illustrative synthetic examples, which contain only a small number of features and a few data points each, so that it is easier to comprehend the obtained results. 
Furthermore, we bench-mark HyperLAU on a dataset, that has already been used to demonstrate HyperTraPS-CT \cite{aga_hypertraps-ct_2024}, both on the complete dataset and after randomly inserting artificial uncertain features. Details and results for both examples can be found in the SI.  We also show that HyperLAU's ability to handle uncertain data increases its performance on test data compared to HyperHMM \cite{moen_hyperhmm_2023} and HyperTraPS \cite{aga_hypertraps-ct_2024}, which are both challenged by uncertainty.

\subsection{HyperLAU on tuberculosis drug resistance data with artificially introduced uncertainties}
\label{subsec art tub}

Additionally to the toy examples, we also want to demonstrate the power of HyperLAU on a real dataset of multidrug-resistance data in tuberculosis. The dataset we use consists of the binary indication of susceptibility or resistance to 10 drugs for each of 295 isolates and has previously been used to test several other accumulation models \cite{moen_hyperhmm_2023,greenbury_hypertraps_2020}. It is a subset of the data originally published in \cite{casali_evolution_2014}.  We used the same ancestral reconstruction of the phenotypic resistance features as \cite{moen_hyperhmm_2023}. This dataset only contains samples with no uncertainties, every feature in every sample can be clearly represented as a zero or a one. To test HyperLAU's ability to reproduce inferences in the face of uncertain data, we artificially introduced  uncertainties in this dataset, by exchanging every character in the original dataset with a possibility of 50\% with a `?', both for the ancestor and the descendant state. The considered features represent resistances against the first-line drugs 1, isoniazid (INH); 2, rifamycin (RIF); 3, pyrazinamide (PZA); 4, ethambutol (EMB) and 5, streptomycin (STR), as well as the second-line drugs 6, amikacin (AMI); 7, capreomycin (CAP); 8, moxifloxacin (MOX); 9, ofloxacin (OFL) and 10, prothionamide (PRO). Training HyperLAU on both datasets, we can compare the learned results (see Figure \ref{tb}).\\

One can see that even with around 50\% uncertain positions in the data, the edges with the highest predicted flux for the original dataset are still identified as such by HyperLAU. In the original dataset there is a clear indication that resistance to either INH or RIF is obtained first, although the order is not unambiguously clear. Afterwards, the resistances against STR, EMB and PZA are acquired, before subsequent steps become more heterogeneous and split into several competitive pathways. These results match the results HyperHMM predicted on the same dataset \cite{moen_hyperhmm_2023}.\\

Considering the results of the data with 50\% inserted uncertainty, the dynamic in the beginning of obtaining resistance first against INH, then against RIF, STR, EMB and PZA is still very clear. However, in this case the version with INH before RIF is clearly favored. It is also visible that several other edges have a higher flux as before, although with a higher CV. This increased uncertainty in inference output is not unexpected given associated uncertainty in the data input. Taken together, despite the challenges involved in artificially obscuring half of the original data, HyperLAU can clearly capture the most likely and outstanding pathways of a real-world data set.
\subsection{Case study 1: Uncertainty in resistance classifications in tuberculosis multi-drug resistance evolution}

To demonstrate how HyperLAU's capacity for uncertain data allows an expansion beyond previous analysis methods, we applied HyperLAU to the full original tuberculosis dataset from \cite{casali_evolution_2014}, including samples for which is no information for some features available (due to the computational difficulty of inferring drug resistances from genome data). These uncertain samples have been excluded in our previous example, and also in the applications of other methods like HyperHMM and HyperTraPS before \cite{moen_hyperhmm_2023,aga_hypertraps-ct_2024}. With HyperLAU it is now possible to take the full set of 1000 samples into account (see Methods for a discussion of how phylogenetically independent transitions were obtained), and get a more complete evolutionary picture than before. \\
\begin{figure}
	\centering
	\includegraphics[scale=0.5]{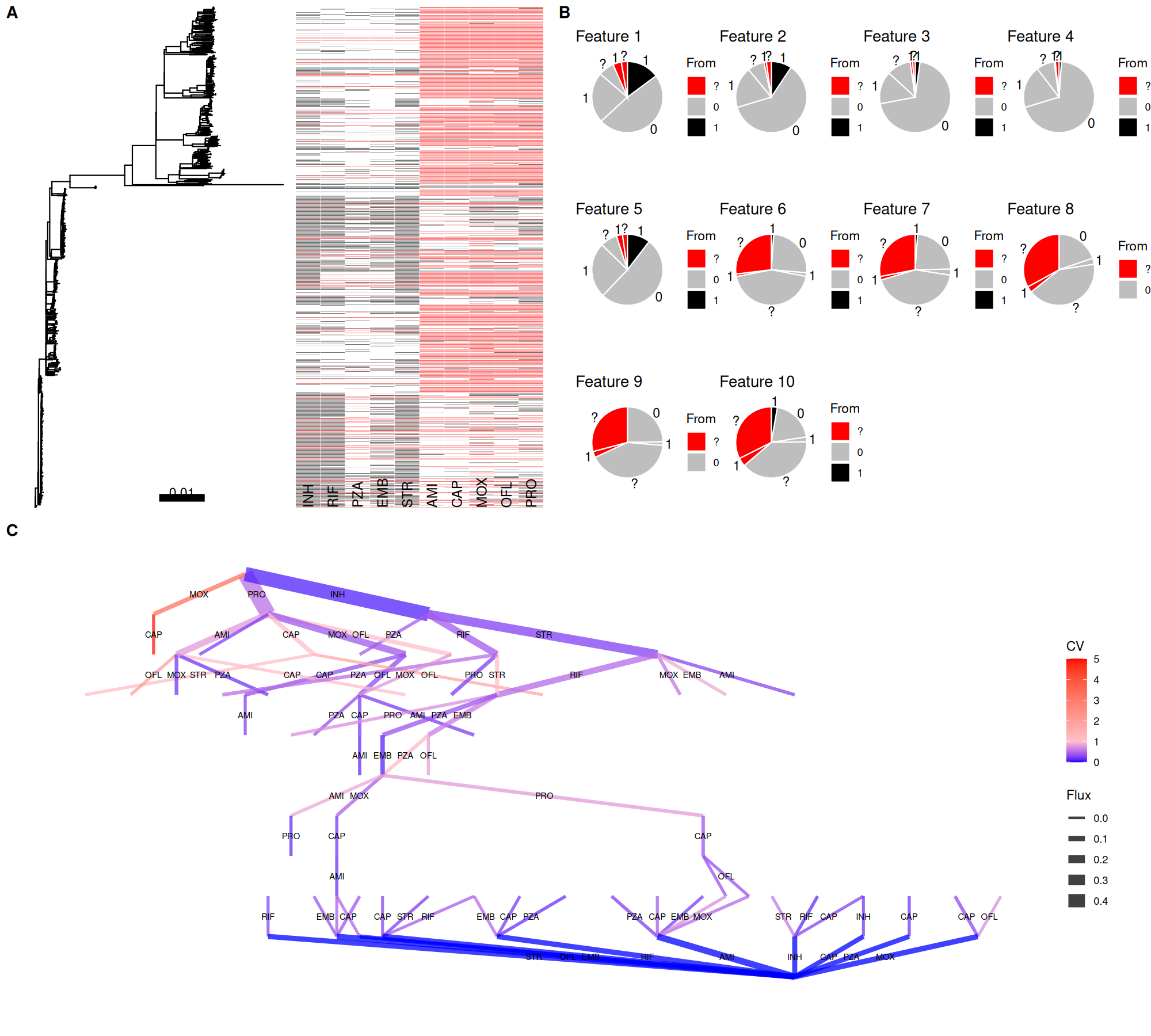}
	\caption{\small \textbf{Inference of anti-microbial evolution in tuberculosis based on a full dataset including uncertainties. (A)} Visualisation of the used dataset from \cite{casali_evolution_2014} embedded in a phylogeny. Each row in the matrix corresponds to a bacterial isolate that is a tip in the phylogeny. Each column in the matrix describes resistance to a different drug: red fields in the profile represent missing data, white fields indicate the absence of resistance, black fields indicate the presence of resistance. \textbf{(B)} Illustration of the transitions (not the individual isolate profiles) present in this dataset. For each drug, pie segment colour describes the `before' state of that drug in a transition, and the labels on the circumference describe the `after' state. The size of each pie section gives the proportion of transitions for that before-after combination. \textbf{(C)} Transition network visualisation of the fluxes learned by HyperLAU through the evolutionary state space from 000... (top) to 111... (bottom) (as in Figure \ref{graph_abstr}D). The thickness of the edges represents the strength of the flux. The coefficient of variation (CV) is illustrated by the colour. Edges are labelled by the drug resistance that is gained in that step. Only edges with a flux $>0.01$ are shown. Illustration of the bootstrap uncertainty for the first steps is given in Figure \ref{tb_features}.}
	\label{results_independent}
\end{figure}

Figure \ref{results_independent} shows the resistance profiles of the samples embedded in a phylogeny (Figure \ref{results_independent}A) and a visualisation of the transitions occurring in the used input dataset (Figure \ref{results_independent}B), along with a visualisation of the inferred evolutionary pathways obtained by HyperLAU (Figure \ref{results_independent}C).\\

In the visualisation of the results we can see a strong preference for a resistance against INH or PRO as the first one to be obtained. This clear preference of INH compared to RIF is especially interesting because existing tools like HyperHMM, that were applied to the reduced dataset without uncertainties, do not highlight such a clear difference between those two features \cite{moen_hyperhmm_2023}. Their results show more the picture of two competing paths towards INH or RIF, as we also have seen in \ref{subsec art tub}. HyperLAU applied to the complete dataset, on the other hand, reports even a high probability for STR to also occur before RIF. Looking at the resistance profiles of the data (Figure \ref{results_independent}A), one can see that many of the data points that show a resistance against STR but not against RIF contain at least one feature without reported data and with that have been excluded in previous studies. However, HyperLAU predicts a high probability that also PRO might be the first resistance, which was not seen in the previous results by other tools, although it should be noted that this result is less certain, what is indicated by a higher CV obtained by the bootstrap resamples (for a more detailed analysis, see SI). Looking at Figure \ref{results_independent}B this seems reasonable, as in Feature 10 (which is represented by PRO) are the transitions ?-? and 0-? the most frequent ones, indicating a lot of uncertainty and little information. This shows that excluding data with incomplete data leads to a certain bias at this point.  After a couple of steps the HyperLAU results loose single highlighted paths and become more diffuse due to the high level of uncertainty. However, the clear highlighting of the INH resistance occurring in the first step as well as the occurrence of STR before RIF, is a significant information that once again shows and emphasizes the important contribution by HyperLAU. 

\subsection{Case study 2: Uncertainty in phenotypic features in $C_4$ photosynthesis evolution}

An early application of EvAM was the remarkable convergent evolution of $C_4$ photosynthesis in plants \cite{williams2013phenotypic}. Involving the acquisition of a collection of physiological, cellular, and gene expression adaptations, $C_4$ has evolved independently over 60 times. Previous work used accumulation modelling to develop a `roadmap' for $C_4$ evolution across these instances, using data from `$C_3-C_4$ intermediates' -- species that have acquired some but not all of the features involved in $C_4$. These results partly explained the remarkably widespread occurrence of $C_4$ by identifying that early steps in its evolution were largely coarse-grained physiological ones that plausibly have positive fitness consequences even in the absence of other $C_4$ traits, `paving the road' for further refinement \cite{williams2013phenotypic}.\\

Due to the technical difficulties of phenotypic observation in diverse $C_3-C_4$ intermediates, much (over 59\%) of the original dataset was incomplete. At the time, this required a naive sampling approach where walkers were simulated in state space and with no guarantee that they would usefully contribute to likelihood estimation, then recorded if they encountered a state compatible with a given incomplete observation. The uncertainty in the resultant roadmap was substantial, and the computational expense was considerable. HyperLAU's natural uncertainty embedding allows this to be addressed much more satisfactorily.\\

\begin{figure}
	\centering
    \includegraphics[width=\textwidth]{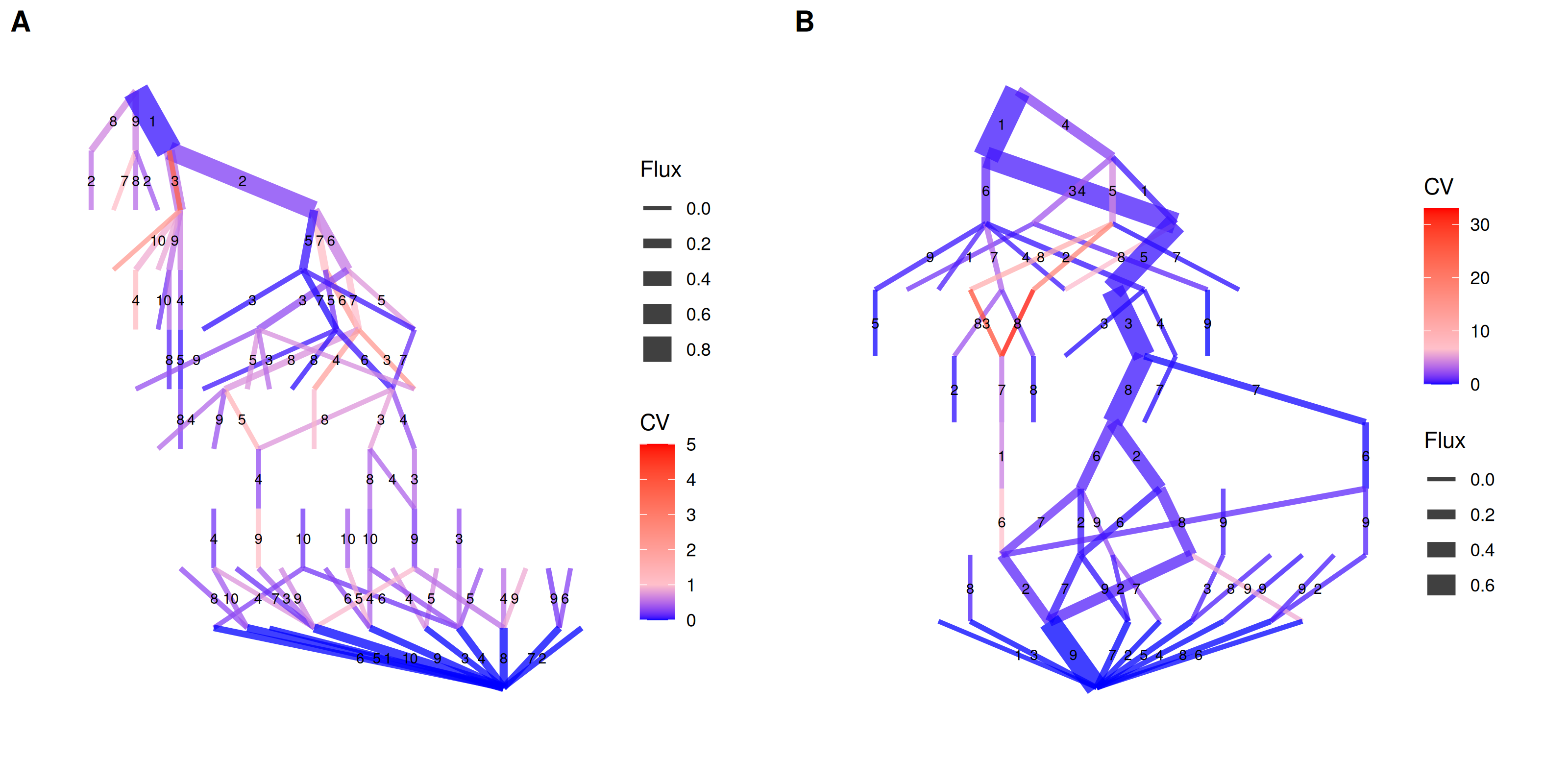}
    \caption{\small \textbf{HyperLAU case studies in photosynthesis and reductive mitochondrial evolution.} Transition networks inferred using HyperLAU, styled as in Fig. \ref{results_independent}C. \textbf{A}: $C_4$ photosynthesis case study, with data from \cite{williams2013phenotypic} (and references therein). Features (where `specificity' means cell-type specificity: (1) large bundle sheath (BS) cells; (2) GDC cell specificity; (3) vein spacing; (4) decarboxylase specificity; (5) PPDK specificity; (6) BS chloroplast number; (7) GDC abundance; (8) PPDK abundance); (9) decarboxylase abundance; (10) RuBisCO abundance. \textbf{B}: Mitochondrial reduction case study, with data from \cite{glastad2025convergent} (and references therein). Features: electron transport chain complexes (1) CI; (2) CII; (3) CIII; (4) CIV; (5) CV; (6) pyruvate dehydrogenase; (7) mtDNA; (8) citric acid cycle steps; (9) iron-sulfur metabolism. \label{other-studies}}
\end{figure}

Fig. \ref{other-studies}A shows the results of HyperLAU inference on the first ten features in this dataset. The first two steps -- enlarged bundle sheath cells and cell specificity of GDC -- remain overwhelmingly favoured as the first evolutionary steps towards $C_4$. Interestingly, the next most likely step (albeit with uncertainty) is predicted to be an increase in the number of bundle sheath chloroplasts. This is a coarse physiological change but was not ranked as an early precursor in the original paper because its ordering posterior was rather bimodal, with density both at early and late acquisition orderings. A similar effect is seen for GDC abundance, here inferred to be earlier than in the original study, and with a bimodal distribution in that study. The explicit handling of uncertainty has here reinforced the scientific message -- that coarse physiological changes, of benefit outside the specific $C_4$ context, are early steps `paving the way' for full $C_4$ evolution.

\subsection{Case study 3: Uncertainty in bioenergetic features in reductive mitochondrial evolution}

A recent study has used EvAM to characterise the pathways of reductive mitochondrial evolution \cite{glastad2025convergent}. Here, in specific (often parasitic) lineages across eukaryotes, mitochondria have lost one or most aspects of their canonical functionality \cite{hampl2024evolutionary, stairs2015diversity, maciszewski2019should}. This reduction to mitochondrion-related organelles or MROs may involve the loss of one or more electron transport chain complexes, the mitochondrial DNA, or the very ability to produce ATP (iron-sulfur cluster metabolism seems to be the most conserved feature of MROs). The study in \cite{glastad2025convergent} showed that the diversity of MROs across eukaryotes can largely be explained by `variations on a theme' of two distinct pathways: the CI-pathway, involving loss of Complex I before other features; and the alt-pathway, involving loss of either CIII, CIV, or steps of the citric acid cycle before other features.\\

However, again, the characterisation of the presence and absence of molecular features of mitochondria in diverse eukaryotes (including many protists) is a challenging task. The \cite{glastad2025convergent} study surveyed the literature to construct a dataset for the EvAM approach. However, several inconsistencies came up across literature sources, with 26 out of 81 organisms containing some conflicting information about presence/absence of features. In the original study this diversity was condensed into two datasets reflecting the `uncertainty means absence' and `uncertainty means presence' limits, and inferences drawn about results that remained consistent across the two. But HyperLAU offers a much more satisfactory approach of treating conflicting observations as uncertain.\\

The results are shown in Fig. \ref{other-studies}B. The structuring in CI- and alt-pathways is still visible, but the CI-pathway is dramatically favoured compared to the alt-pathway in this HyperLAU treatment. The structure of the alt-pathway is more canalised under HyperLAU, with CIV as the most common initial loss step and a previously unidentified mechanism -- loss of CIV, then CI, then following the CI-pathway -- showing some consistent probability. The outcome of accounting for uncertainty in this case has been to reweight the previous inference outcomes, providing a more parsimonious explanation of the diverse dynamics observed across eukaryotes: the highly probable CI-pathway is highly canalised and predictable, and tied to limiting metabolic impact of mitochondrial reduction \cite{glastad2025convergent}.

\section{Discussion}

We have introduced HyperLAU as a new tool to learn and model accumulation processes based on binary data, which is able to deal with precisely specified uncertainties in input data that consists of an ancestor and a descendant state.  An additional advantage of HyperLAU is, that there can be arbitrary interactions between different features taken into account, while there also is the option to reduce the parameter set and follow an approach as used for MHNs or reduced HyperTraPS models. Our different case studies have pointed out the power to use data without the need to exclude samples due to knowledge gaps. Furthermore, they have shown that, given a dataset without any uncertainties, our results align with those of other models.\\

The HyperLAU algorithm assumes the different transitions in the input data to be independent of each other, which may not be the case for uncertain phylogenetic or longitudinal data. We accounted for this dependence by removing non-independent transitions. To include all non-independent transitions in a dataset, it would be necessary to ensure that the interpretation of the uncertainty markers among all strings that represent one and the same position in the phylogeny are consistent. For the moment, HyperLAU is not able to provide this, and further work has to be done. Nevertheless, we believe that the possibility to handle uncertainties in ancestor and descendant of pairs in HyperLAU is the first step in this direction. Our demonstration that this increases HyperLAU's performance over EvAM alternatives supports this picture.\\

Some other aspects of our approach can be further generalised in future. The simulated annealing optimisation is a random process, so evaluating the results for different random seeds will enhance robustness at a cost of additional computational time. We assume (commonly with other evolutionary accumulation approaches) that features are accumulated irreversibly, a simplification that does not necessarily hold true in the actual underlying biological process \cite{renz_evolutionary_2024}. For implementing this reversibility \cite{johnston_hypercubic_2024} in HyperLAU, further work is needed, but we think that our linear algebra approach could also be a reasonable starting point, for example by allowing also upper-diagonal entries in the transition matrix to be positive.\\

An important consideration in applying HyperLAU is that all states compatible with a given observation are considered equally likely. In real contexts it may be, for example, that prior knowledge suggests that a particular feature's prevalence is systematically low (or high) across a dataset, and so states assuming a 0 (or 1) for an uncertainty marker should be considered more likely. Building a prior-probability specification into HyperLAU's approach -- either in the likelihood calculation or via a Bayesian embedding -- would further generalise this picture.\\

Furthermore, HyperLAU in its current form cannot account directly for false negative and false positive errors in the dataset. Uncertainties have to be identified as such before. It might be a valuable further direction to introduce also a component into the matrix-vector multiplications that accounts for this.\\

A disadvantage compared to other models might be the longer run time when considering a larger amount of features. Additionally to the up-scaling due to matrix-vector multiplications with bigger matrices, the run-time also varies with the amount of uncertainty markers in the input data. With more uncertainty, considerably more possible pathways have to be taken into account to calculate the likelihood. A possible way to address this problem could be the usage of another, more efficient optimization method, or the implementation of a tensor format as in \cite{schill_reconstructing_2024}.\\

Compared to HyperTraPS-CT and MHN, which work in a continuous-time setup, HyperLAU is not yet able to take continuous time into account. As shown by a case study in \cite{aga_hypertraps-ct_2024}, modelling of such an accumulation process in continuous time instead of discrete time-steps can lead to subtly different results, when considering less-frequent features that have a fast capture time. Embedding this approach in a continuous-time framework would be an important next step, which further combines the advantages of the different approaches. For now, we hope that this method is a useful contribution towards the goal of learning accumulation processes from broader, uncertain data without a prior selection regarding completeness. 

\section*{Acknowledgments}
This work was supported by the Trond Mohn Foundation [project HyperEvol under grant agreement No. TMS2021TMT09 to IGJ], through the Centre for Antimicrobial Resistance in Western Norway (CAMRIA) [TMS2020TMT11].

\bibliographystyle{apalike}
\bibliography{Paper_Uncertain_states}

\newpage
\section*{Supplementary Information}
\renewcommand{\thefigure}{S\arabic{figure}}
\setcounter{figure}{0}
\renewcommand{\thesection}{S\arabic{section}}
\renewcommand{\thesubsection}{S\arabic{section}.\arabic{subsection}}
\setcounter{section}{1}
\renewcommand{\theequation}{S\arabic{equation}}
\setcounter{equation}{0}

\pagestyle{empty}

\subsection{Calculation of the likelihood function}
\label{SI_likleihood}

For every observation $k^{i},\ i \in \{a,d\}$, in our dataset the information of compatibility is stored in a vector $c^{a}$ for the \textit{ancestor (a)} string and a vector $c^{d}$ for the \textit{descendant (d)} string, where we set $c^{i}_{s} = 1$ if node $s$ is compatible with $k^{i}$ and 0 otherwise. Because we assume that every trajectory starts at $0^{L}$ and in every evolutionary step exactly one new feature is obtained until we end up at $1^{L}$, we need $L+1$ steps to simulate a whole pathway. 

For a given transition matrix $A$ of the hypercube, we can compute a vector $P_{.,t}$ for every evolutionary step $t=0,...,L$, where $L$ is the length of the strings, i.e. the number of considered features. $P_{s,t}$ is then the probability of a trajectory being in evolutionary step $t$ at node $s$, where $s$ is the natural number that corresponds to the binary string. That means every $P_{.,t}$ gives a probability distribution of occupancy for the different nodes of the hypercube at step $t$. First we set $P_{.,0} = [1,0,0,...]$. This reflects our assumption that every trajectory has to start in the state 000..., where none of the features have yet been acquired. Then

\begin{equation}
	\label{matrix_vector_prob_supp}
	P_{.,t} = A\cdot P_{.,t-1} = A^{t} \cdot P_{.,0}.
\end{equation}

We can now go step-wise through the columns of the matrix $P = (P_{.,t})_{0\leq t\leq L}$, i.e. the evolutionary steps, and determine the likelihood assuming the ancestor data has its origin in this step. To keep only the information about the probability of being in a state that is compatible with an ancestor in our dataset, we multiply $P_{.,t}$ entry-wise (denoted $\star$) with $c^{a}$. For every entry-wise product vector $c^{a}\star P_{.,t}$ where there is at least one entry that is bigger than zero, we assume the ancestor can originate from step $t$. Therefore we continue by applying the same matrix-vector approach to get vectors $Q_{.,t'}^{t,a}, \ t\leq t' \leq L$ containing the probability distribution for being in a certain node of the hypercube at step $s$ given that the trajectory reached already the ancestor state in step $t$ before.  For $t'=t$, $Q_{.,t'}^{t,a}$ is set to the vector that results when multiplying $P_{.,t}$ entry-wise with $c^{a}$:
\begin{equation*}
	Q_{.,t'}^{t,a} = c^{a}\star P_{.,t} \text{ for } t'=t.
\end{equation*} From here we continue just by multiplying with the transition matrix in the same way as in (\ref{matrix_vector_prob_supp}):
\begin{equation*}
	Q_{.,t'+1}^{t,a} = A\cdot Q_{.,t'}^{t,a}\quad \text{ for } t'={t,...,L}.
\end{equation*}
Then we have to multiply every $Q_{.,t'}^{t,a}$ this time with $c^{d}$ so that we end up with the sum of the probabilities also compatible with the descendant state of the data point. This is exactly the probability that a trajectory first goes through a node that is compatible with the ancestor state of the dataset, and afterwards through a node that is compatible with the descendant state. We specifically allow the ancestor and the descendant state to be the same one. \\
Summarized, the likelihood function of seeing a data point can be described by the following formula:
\begin{equation}
	\label{lik_one_point_supp}
	\mathcal{L}((k^{a},k^{d})|A) =\frac{2}{(L+1)(L+2)}\cdot \sum_{t=0}^{L}\sum_{t'=t}^{L} c^{d}\cdot Q_{.,t'}^{t,a}
\end{equation}
with
\begin{equation*}
	Q_{.,t'}^{t,a} = A^{t'-t}\cdot \left(c^{a}\star P_{.,t}\right).
\end{equation*}
Here the factor $2/(L+1)(L+2)$ is the normalization constant reflecting the $(L+1)(L+2)/2$ possible combinations of nodes on the hypercube at which ancestor and descendant can be sampled.

Calculating Equation (\ref{lik_one_point_supp}) for every entry in our dataset, taking the logarithm, and summing up the obtained log-likelihoods, we get an overall logarithmic likelihood of obtaining the data in the dataset, given the transition matrix. This is also the function that needs to be optimized by adapting the transition or rate matrix.

\begin{algorithm}
	\caption{Pseudocode of the calculation of the likelihood with the HyperLAU method.}\label{likelhood_code}
	\textbf{Input:} Vectors with ancestor and descendant states of the dataset and the corresponding frequency, length of the strings $L$, transition matrix $A$
	
	\textbf{Output:} log-likelihood of seeing the input data given the transition matrix
	
	\begin{algorithmic}[1]
		\State Set $P_{.,0} = (1,0,...,0)$ (vector of length $2^L$, indicates probability distribution at $t=0$)
		\For { $t$ in $1:L$}
		\State $P_{.,t} = A \cdot P_{.,t-1}$ 
		\EndFor
		\For {every point in the dataset}
		\State vector (length $2^L$) $c^a$ = binary entries, indicate compatible states for the ancestor
		\State vector (length $2^L$) $c^d$ = binary entries, indicate compatible states for the descendant
		\For {$t$ in $0:L$}
		\State $P_{.,0}^{\text{comp}}= c^a \star P_{.,t}$ with $\star$ entry-wise multiplication
		\If {sum($P_{.,0}^{\text{comp}}$) $\neq 0$}
		\State Initialize $Q$ = null-matrix of dimension $2^L \times L$
		\State $Q_{.,t}$ = $P_{.,0}^{\text{comp}}$
		\For {$t'$ in $t+1:L$}
		\State $Q_{.,t'}$ = $A \cdot Q_{.,t'-1}$
		\EndFor
		\For {$t'$ in $0:L$}
		\State $Q_{.,s}^{\text{comp}} = c^d \star Q_{.,t'}$
		\If {sum($Q_{.,t'}^{\text{comp}}$) $\neq 0$}
		\State store sum($Q_{.,t'}^{\text{comp}}$)
		\EndIf
		\EndFor
		\EndIf
		\EndFor
		\State $\mathcal{L}' = $ sum of all stored sum($Q_{.,t'}^{\text{comp}}$) multiplied with normalization factor
		\State $\mathcal{L} = \mathcal{L'}$ multiplies with the corresponding frequency of the data point
		\State Store the logarithm of $\mathcal{L}$ as the log-likelihood for this transition
		\EndFor
		\State Sum over log-likelihoods for all data points
	\end{algorithmic}
\end{algorithm}

\subsection{Optimisation and convergence}
\label{optimisation_convergence}
The optimization process follows a simulated annealing process where in every loop the entries of the transition matrix, or in the reduced case the rate matrix, are changed a small amount. The reason why we preferred simulated annealing over a gradient based approach is its ability to escape local optima and find the global optimum. This is especially important because we have no prior knowledge about the shape of the optimization landscape.

The initial temperature is set to 1 and decreased in every loop by dividing by a certain factor, which can be specified as an input parameter. For all results here presented, the temperature was divided by 1.001 in every step. As soon as the temperature falls below a threshold of $10^{-6}$, the iterations and with that the optimization process are stopped. This choice of parameter values gave us a sufficient level of convergence for the datasets we considered (for the progression trajectories of the likelihood see SI). All results of the tuberculosis, photosynthesis, and mitochondrial evolution case studies are obtained under model $F$.

With each iteration, every parameter is adjusted individually by a random number from the uniform distribution between -0.025 and 0.025. If this leads to a negative value, the parameter is set to zero. The new parameter matrix is accepted over the current one, if the corresponding log-likelihood is bigger than the one before, or if 
\begin{equation*}
	\exp\left(\frac{-(l_{\text{old}}-l_{\text{new}})}{T}\right) > U(0,1),
\end{equation*}
where $l_{\text{old}}$ and $l_{\text{new}}$ are the two log-likelihoods, and $T$ the current temperature.

As initial guess of the transition matrix for model $F$, we implemented the matrix that represents a uniform distribution of edge weights among all possible nodes that can be directly achieved from the current state.  \\
For models 1-4 we have to infer the parameters contained in a rate matrix. Here we set only the base rates to one and all rates describing influences by sets of already obtained features to zero. Different than in the case of the transition matrix, here are all entries part of the optimization process and can be changed during the Simulated Annealing. These initial guesses are a natural choice when there are no prior information or assumptions available. However, it is also possible to implement another initial guess, if necessary. 

\subsection{Manipulation of the datasets to artificially insert uncertainty.} 
\label{including_uncertainty}
To establish that our model is able to predict the most likely evolutionary pathways even with uncertain states in the data, we created some test datasets by inserting random uncertainty markers in the original ones. For the toy-example case, we included the uncertainty feature-wise. For the tuberculosis dataset, we spread them uniformly over all positions. For doing so, for each relevant position in every string a random number was drawn from the uniform distribution between zero and one. If the random number was smaller than a specified threshold, the corresponding position in the dataset was changed to a `?', if not, the original binary entry was kept. This procedure was done with thresholds 0.4 for the toy-example and 0.5 for the tuberculosis data. Doing so we obtained versions of the toy-example dataset with around 40\% uncertainty markers in every feature each, as well as a version of the tuberculosis data with uncertainty markers in around 50\% of all positions. 

\subsection{Implementation.} The code is implemented in C++ and is freely available on Github: \url{https://github.com/JessicaRenz/HyperLAU} and has DOI 10.5281/zenodo.23598345. The Armadillo library for linear algebra and scientific computing \cite{sanderson_armadillo_2016,sanderson_practical_2019} is used in this code. Additionally there are also R scripts for creating the visualisations of the output, string manipulation and tree processing, which use the libraries ggplot2 \cite{wickham_ggplot2_2016}, stringr \cite{wickham_stringr_2023}, ggraph \cite{pedersen_ggraph_2024}, ggpubr \cite{kassambara_ggpubr_2023}, igraph \cite{csardi_igraph_2006,csardi_igraph_2024}, ggrepel \cite{slowikowski_ggrepel_2024}, readxl \cite{wickham_readxl_2025}, tidyr \cite{wickham_tidyr_2024} and dplyr \cite{wickham_dplyr_2023}.

\subsection{HyperLAU on toy examples}
\begin{figure}
	\centering
	\includegraphics[scale = 0.5]{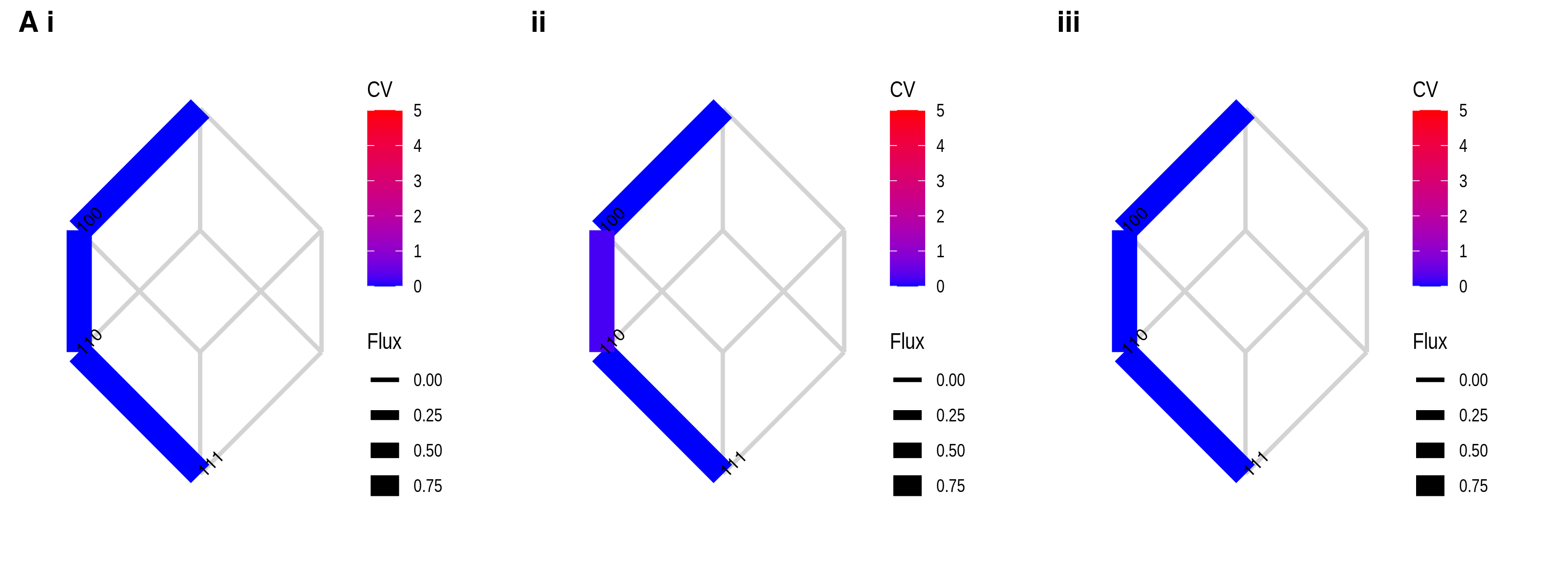}\\
	\includegraphics[scale=0.5]{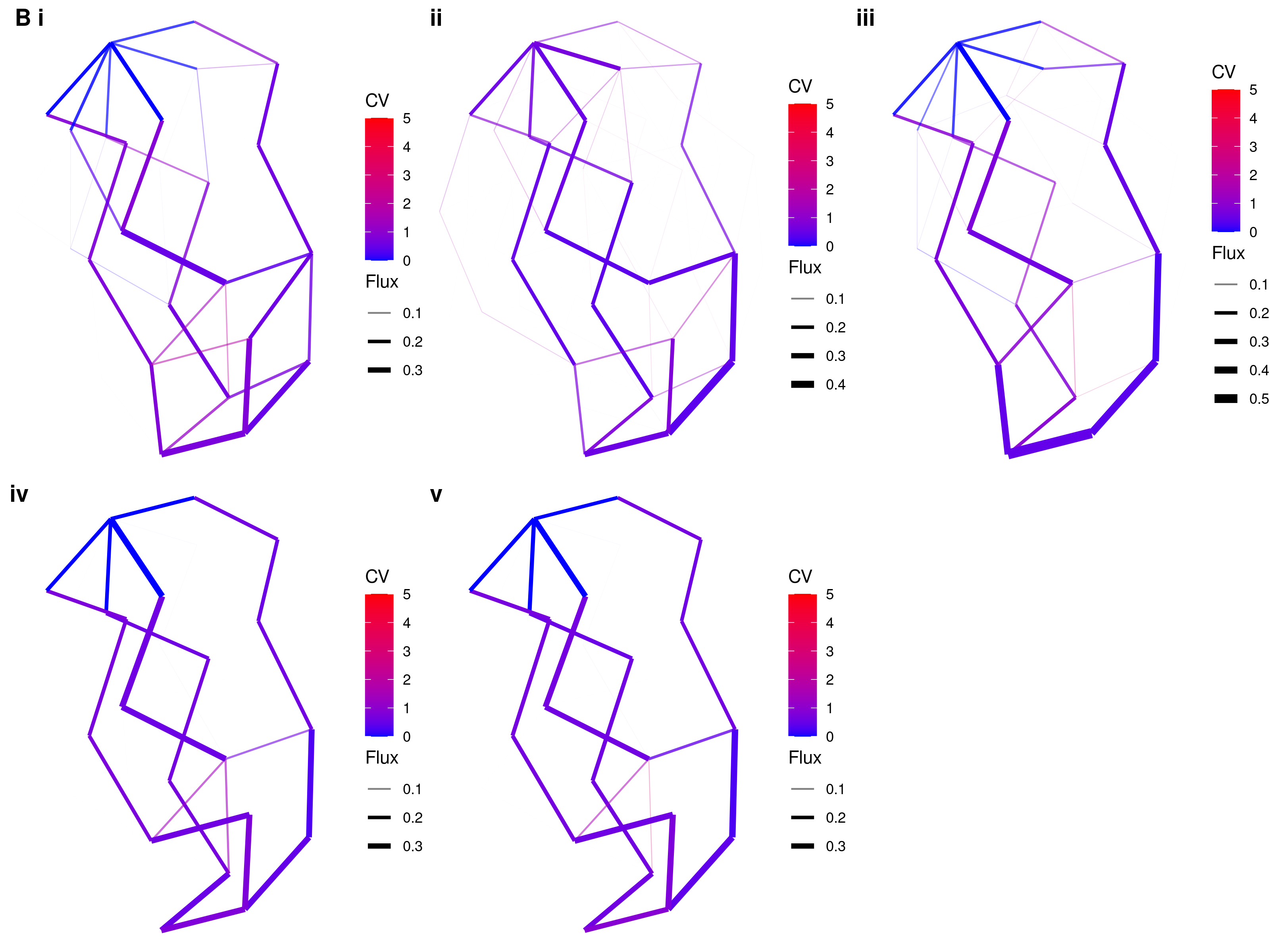}
	\caption{\small \textbf{Visualization of evolutionary pathways inferred by HyperLAU based on some toy examples.} Plots show inferred transition networks through the evolutionary state space from 000... (top) to 111... (bottom) (as in Figure \ref{graph_abstr}D). In all plots, the thickness of the edges represents the probability flux between the corresponding state nodes (all coloured edges have minimum 0.05). Coefficient of variation (CV) is illustrated by the colour. \textbf{A}: Toy dataset including the data points 0?0 - ?00, 10? - 1?0 and 11? - ?11, under model $F$ (i), model 1 (ii) and model 2 (iii), showing a clear path 000-100-110-111 supported in all cases. \textbf{B}: Artificially generated data, where pairs of features influence the occurrence of others (as used in \cite{aga_hypertraps-ct_2024}). HyperLAU model $F$ (i) reproduces established evolutionary pathways from HyperTraPS-CT \cite{aga_hypertraps-ct_2024} (ii). The structure of these inferred pathways under model $F$ remains robust with around 40\% of the data for a particular feature are made uncertain (iii). Inference using model 2 (more restricted, pairwise interaction) is forced to approximate the higher-order true dynamics (iv), and also remains robust when 40\% of the data for a particular feature are made uncertain (v). Results from other artificial-obscuring protocols are shown in Figures \ref{other_features_model-1} and \ref{other_features_model2}. The likelihood traces throughout the optimisation processes are shown in Figures \ref{lik_toy1}-\ref{lik_features_model2}. }
	\label{res_toy_ex}
\end{figure}

First we demonstrate the effect of our algorithm on illustrative synthetic examples, which contain only a small number of features and a few data points each, so that it is easier to comprehend the obtained results (see concrete calculations in Subsection \ref{calc_toy}). The first toy example is specified by the longitudinal dataset containing the following three entries: 0?0 - ?00, 10? - 1?0 and 11? - ?11. When considering only three features, there can be no interactions of higher dimension than two, so it is sufficient to consider the models $F$, 1 and 2. The results given by the different models are illustrated in Figure \ref{res_toy_ex}A.

To indicate the pairings compatible with the input data, we have to take into account that every `?' could either be a zero or a one. From the first data point, we therefore get the compatible transitions 000 - 000 and 000 - 100. The other combinations, having 010 in the ancestor state, are not possible, because we assume the attainment of a feature to be irreversible, and the second position in the descendant is clearly marked by a zero. Following the same logic, the second sample gives us either 100 - 100 or 100 - 110, and the last one 110 - 111 or 111 - 111. The combination in which ancestor and descendant are represented by the same string, can't tell us much about the further progress from the corresponding state. That restricts us to the set of  000 - 100, 100 - 110 and 110 - 111, which depicts the clear path 000 - 100 - 110 - 111. Looking at Figure \ref{res_toy_ex}A, this is predicted by all three models. The explicit calculation for this example is given in Subsection \ref{calc_toy}. 

The other toy example we present here has been used to demonstrate the continuous-time version of HyperTraPS \cite{aga_hypertraps-ct_2024}. The data are generated by a process where pairs of already obtained features influence the further  evolutionary trajectory in a non-additive way beyond the pairwise interaction picture of MHN. In Figure \ref{res_toy_ex}B, we demonstrate that the main tendency of the most likely pathways learned under the full model $F$ allowing arbitrary interactions  fit the ones that are learned by HyperTraPS-CT in \cite{aga_hypertraps-ct_2024} (see Figure \ref{res_toy_ex}B i - ii). 

\begin{figure}[h]
	\centering
	\includegraphics[scale = 0.5]{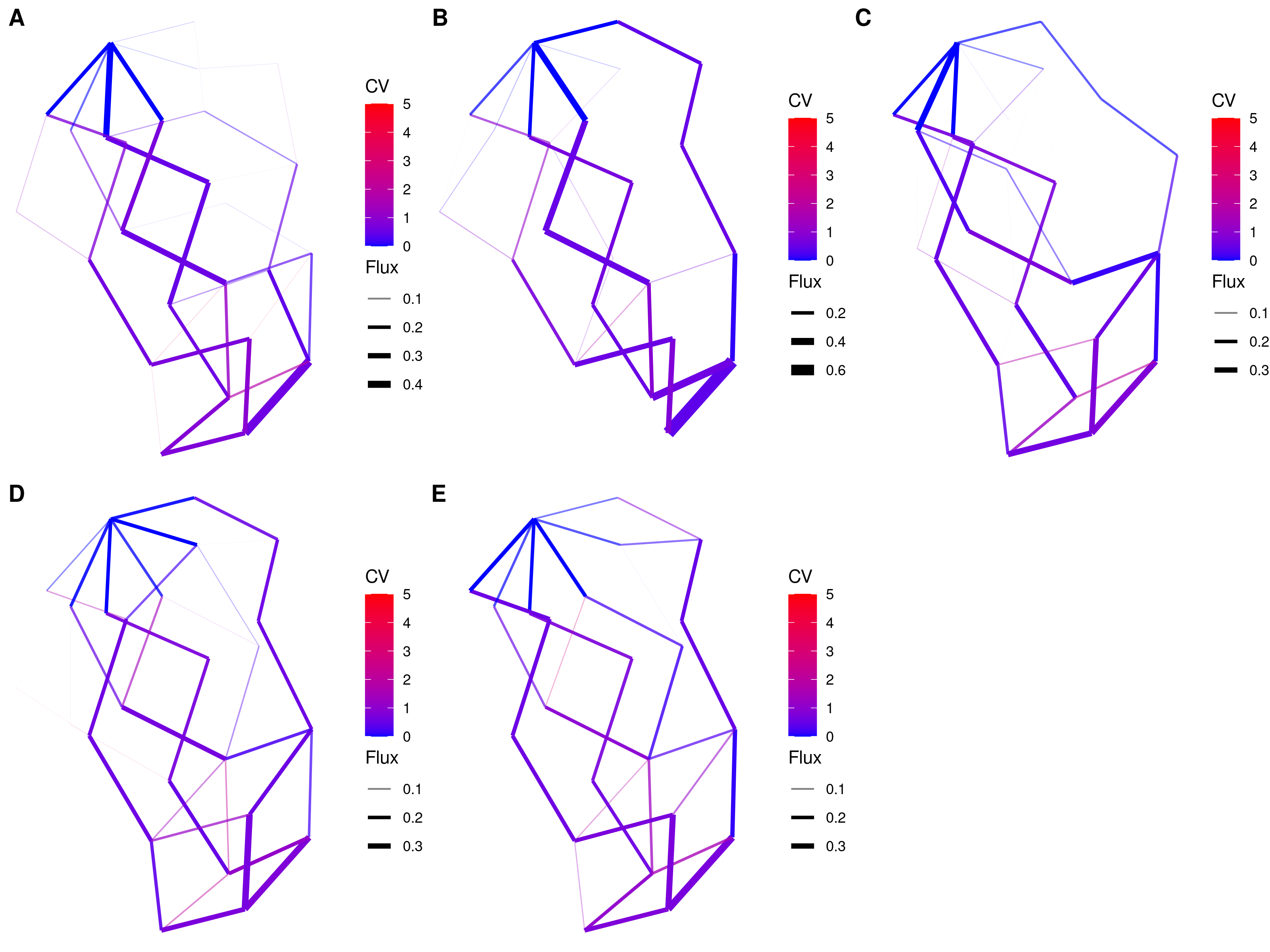}\\
	\caption{\small \textbf{Inferred pathways for the second dataset used in section 3.1 under model $F$, but with approximately 40\% randomly inserted uncertainty markers} in \textbf{A}: feature 2, \textbf{B}: feature 3, \textbf{C}: feature 4, \textbf{D}: feature 5, \textbf{E}: feature 6. The strength of the flux is indicated by the edge-width and the CV is marked by the colour of the edges.   }
	\label{other_features_model-1}
\end{figure}

\begin{figure}[h]
	\centering
	\includegraphics[scale = 0.5]{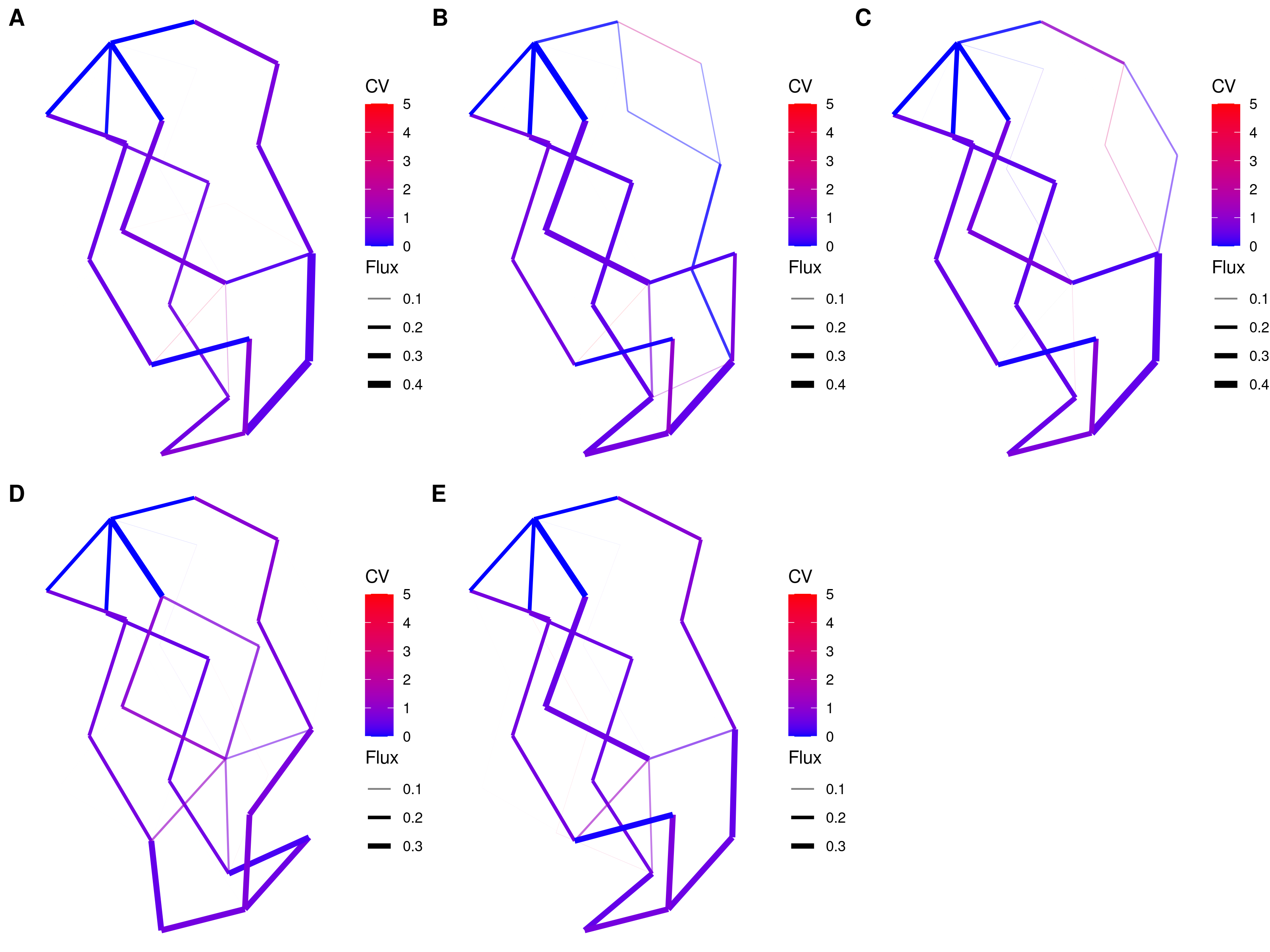}\\
	\caption{\small \textbf{Inferred pathways for the second dataset used in section 3.1 under model 2, but with approximately 40\% randomly inserted uncertainty markers} in \textbf{A}: feature 1, \textbf{B}: feature 2, \textbf{C}: feature 4, \textbf{D}: feature 5, \textbf{E}: feature 6. The strength of the flux is indicated by the edge-width and the CV is marked by the colour of the edges.   }
	\label{other_features_model2}
\end{figure}

Also, after randomly replacing around 40\% of the zeroes and ones representing feature 1 by uncertainty markers, the edges having a high flux are almost the same (Figure \ref{res_toy_ex}B iii). Model 2, which only takes into account the influence of existing individual factors, is not able to capture the same full structure (Figure \ref{res_toy_ex}B iv). Here we see again, that it can be necessary to allow higher-order interactions to get a better picture of the true dynamics. But also under model 2 the inclusion of around 40\% uncertainty in, for example, feature 3 (Figure \ref{res_toy_ex}B v) leads to a result that is extremely close to the results of the original data. This also holds for datasets including uncertainty in some of the other features (see Figures \ref{other_features_model-1} and \ref{other_features_model2}) and demonstrates that HyperLAU can catch the underlying dynamics, even when up to 40\% of the information about one feature is missing. This scenario is common, for example, when datasets are merged from several sources with a different set of features reported. 

\subsection{Calculation of the likelihood for a toy example}
\label{calc_toy}

This illustrative dataset consists of the following data points: 
\begin{enumerate}
	\item 0?0 ?00
	\item 10? 1?0
	\item 11? ?11
\end{enumerate}
We can now calculate for every of these three data points the corresponding vectors $c^{a}$ and $c^{d}$ that indicate the compatible states in the hypercube:
\begin{enumerate}
	\item $c^{a} = (1,0,1,0,0,0,0,0), \quad c^{d} = (1,0,0,0,1,0,0,0)$
	\item $c^{a} = (0,0,0,0,1,1,0,0), \quad c^{d} = (0,0,0,0,1,0,1,0)$
	\item $c^{a} = (0,0,0,0,0,0,1,1), \quad c^{d} = (0,0,0,1,0,0,0,1)$.
\end{enumerate}

By the assumption, that every trajectory starts at $(0,0,0)$ at $t=0$, the probability distribution $P_{.,0}$ has to be

\begin{equation*}
	P_{.,0} = (1,0,0,0,0,0,0,0).
\end{equation*}

$P_{.,1}, \ P_{.,2}$ and $ P_{.,3}$ can be obtained by multiplication with the transition matrix $A = (a_{i,j})$, where $a_{i,j}$ is the probability of the transition from node $j$ to node $i$. So we get

\begin{equation*}
	\begin{split}
	P_{.,1} &= (0, a_{1,0}, a_{2,0},0,a_{4,0},0,0,0), \\
	 P_{.,2} &= (0,0,0,a_{2,0}a_{3,2} + a_{1,0}a_{3,1},0,a_{4,0}a_{5,4} + a_{1,0}a_{5,1},a_{4,0}a_{6,4}+a_{2,0}a_{6,2},0)\text{ and }\\
	 P_{.,3} & = (0,0,0,0,0,0,0,1).
	\end{split}
\end{equation*}

Now we can calculate the likelihood for data point 1. First, we have to multiply the vectors $P_{.,i}$ coordinate-wise with $c^{a}$ to keep only the information that fit our data point. 

\begin{equation*}
	\begin{split}
		c^{a} \star P_{.,0} &= (1,0,0,0,0,0,0,0)\\
		c^{a} \star P_{.,1} &= (0,0,a_{2,0},0,0,0,0,0)\\
		c^{a} \star P_{.,2} &= (0,0,0,0,0,0,0,0)\\
		c^{a} \star P_{.,3} &= (0,0,0,0,0,0,0,0).
	\end{split}
\end{equation*}

In a next step we assume now that the ancestor state originates from step $t=0$. In this case we have to set $Q_{.,0} = c^{a}\star P_{.,0} = P_{.,0}$ and with that we obtain $Q_{.,1} = P_{.,1},\ Q_{.,2} = P_{.,2}$ and $Q_{.,3} = P_{.,3}$. Now we just have to check the compatibility with the descendant state by multiplying $Q_{.,i}$ entry-wise with $c^{d}$ and obtain

\begin{equation*}
	\begin{split}
		c^{d}\star Q_{.,0} &= (1,0,0,0,0,0,0,0)\\
		c^{d}\star Q_{.,1} &= (0,0,0,0,a_{4,0},0,0,0)\\
		c^{d}\star Q_{.,2} &= (0,0,0,0,0,0,0,0)\\
		c^{d}\star Q_{.,3} &= (0,0,0,0,0,0,0,0).
	\end{split}
\end{equation*}

The same procedure we have to follow with assuming that the ancestor state originates from step $t=1$. In this case we set $Q_{.,1} = c^{a}\star P_{.,1} = (0,0,a_{2,0},0,0,0,0,0)$ and obtain by multiplying with $A$:

\begin{equation*}
	\begin{split}
		Q_{.,2} & = (0,0,0,a_{2,0}a_{3,2},0,0,a_{2,0}a_{6,3},0)\\
		Q_{.,3} &= (0,0,0,0,0,0,0,a_{2,0}a_{3,2} + a_{2,0}a_{6,3}).
	\end{split}
\end{equation*}

Checking compatibility with the descendant state gives:

\begin{equation*}
		\begin{split}
			c^{d}\star Q_{.,1} &= (0,0,0,0,0,0,0,0)\\
			c^{d}\star Q_{.,2} &= (0,0,0,0,0,0,0,0)\\
			c^{d}\star Q_{.,3} & = (0,0,0,0,0,0,0,0),
		\end{split}
\end{equation*}

which means that the assumption that the ancestor was sampled at $t=1$ cannot be true. The fact that both $c^{a}\star P_{.,2}$ and $c^{a}\star P_{.,3}$ are the null vectors indicate that the probability of the ancestor data point originate from step $t=2$ or $t=3$ is zero and we therefore don't need to consider these option any further. 

Taking together the positive entries in all $c^{d}\star Q_{.,i}$ vectors and normalize them, we obtain the following likelihood of seeing the first data point given transition matrix $A$:

\begin{equation*}
	\mathcal{L}((0?0,?00) | A) = \frac{2}{4\cdot 5} (1+a_{4,0}) = \frac{1}{10} (1+a_{4,0}).
\end{equation*}

Following the same procedure for 2. and 3., we get the likelihoods

\begin{equation*}
	\mathcal{L}((10?,1?0)|A) = \frac{1}{10}(a_{4,0}+a_{4,0}a_{6,4})
\end{equation*}

and

\begin{equation*}
	\mathcal{L}((11?,?11)|A) = \frac{1}{10}(a_{4,0}a_{6,4} + a_{2,0}a_{6,2} + 1).
\end{equation*}

Taking the logarithm of the product of the likelihood expressions for the different data points, one ends up with the final log-likelihood for seeing this dataset given a matrix $A$ of

\begin{equation*}
	\begin{split}
	&\log\left(\frac{1}{10}(1+a_{4,0})\right) + \log\left(\frac{1}{10}(a_{4,0}+a_{4,0}a_{6,4})\right) + \log\left(\frac{1}{10}(a_{4,0}a_{6,4} + a_{2,0}a_{6,2} + 1)\right)\\
	=& \log\left(\left(\frac{1}{10}\right)^{3}(1+a_{4,0})(a_{4,0}+a_{4,0}a_{6,4})(a_{4,0}a_{6,4} + a_{2,0}a_{6,2} + 1)\right).
	\end{split}
\end{equation*}

\subsection{Model comparison with HyperTraPS and HyperHMM}
\label{subsec_analysis_comparison}

In this subsection, we provide a more detailed comparison of HyperLAU and some comparable models, namely HyperHMM \cite{moen_hyperhmm_2023} and HyperTraPS \cite{aga_hypertraps-ct_2024}. For doing so, we took the tuberculosis dataset \cite{casali_evolution_2014}, we already considered in case study 1. There, we restricted ourselves to the datapoints without uncertainties and created different sub-datasets containing data with different string lengths $L=3,...,10$.  In each of these subsets, from each datapoint the first $L$ features are considered. Then, every dataset was randomly split into a training set and a test set. From the training set, we created additional datasets by inserting different proportions of artificially inserted questionmarks. \\
Then, all three models were trained on all datasets. For running HyperHMM on the datasets containing questionmarks, it was necessary to ignore any datapoint containing at least one uncertainty marker. For HyperTraPS, only the datapoints with a fully available ancestor string could be taken into account. \\

\begin{figure}
	\centering
	\includegraphics[scale = 0.5]{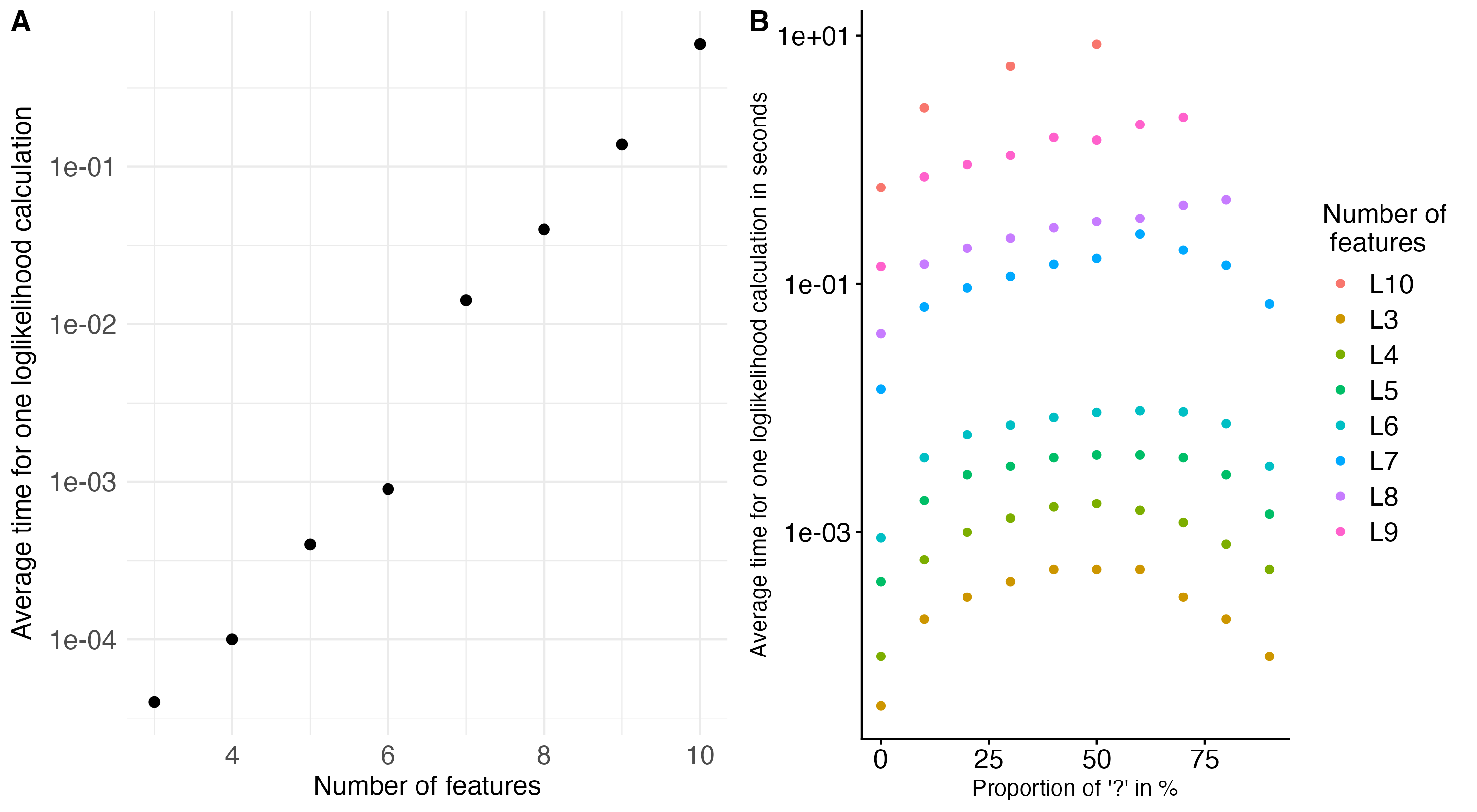}
	\caption{The behaviour of average time for one calculation of a log-likelihood In HyperLAU, (a) depending on the number of features in datasets without any uncertainty markers, (b) depending on the proportion of uncertainty markers in the input dataset. The y-axes are in logarithmic scale.} 
	\label{runtime_plots}
\end{figure}

The average runtime for one log-likelihood calculation within the simulated annealing process in HyperLAU grows exponentially in the number of considered features (see Figure \ref{runtime_plots}(a)). An increase of the proportion of uncertainty markers first increases also the runtime (although slower than exponential). However from approximately 50\% of uncertain features in the input data, the runtime is decreasing again when $L\leq 7$ (see Figure \ref{runtime_plots}(b)). Compared to the other two models, however, HyperLAU's runtime is significantly inferior. While both HyperTraPS and HyperHMM finish the inference for the dataset with string length $L=10$ within seconds (HyperTraPS: 19s, HyperHMM: 5s), HyperLAU needs approximately 7.5 hours (without bootstrap resamples).\\

The strength of HyperLAU lies in the fact that the main path can be inferred for almost any number of features, even with high levels of uncertainty. By contrast, the fact that data points that contain one or more uncertainty markers cannot be used for inference with HyperTraPS or HyperLAU and the number of datapoints in the input data drastically decreases as the proportions of questionmarks increases, those models have more problems to infer the main paths that were originally found by all three models.\\

\begin{figure}
	\centering
	\includegraphics[scale = 0.6]{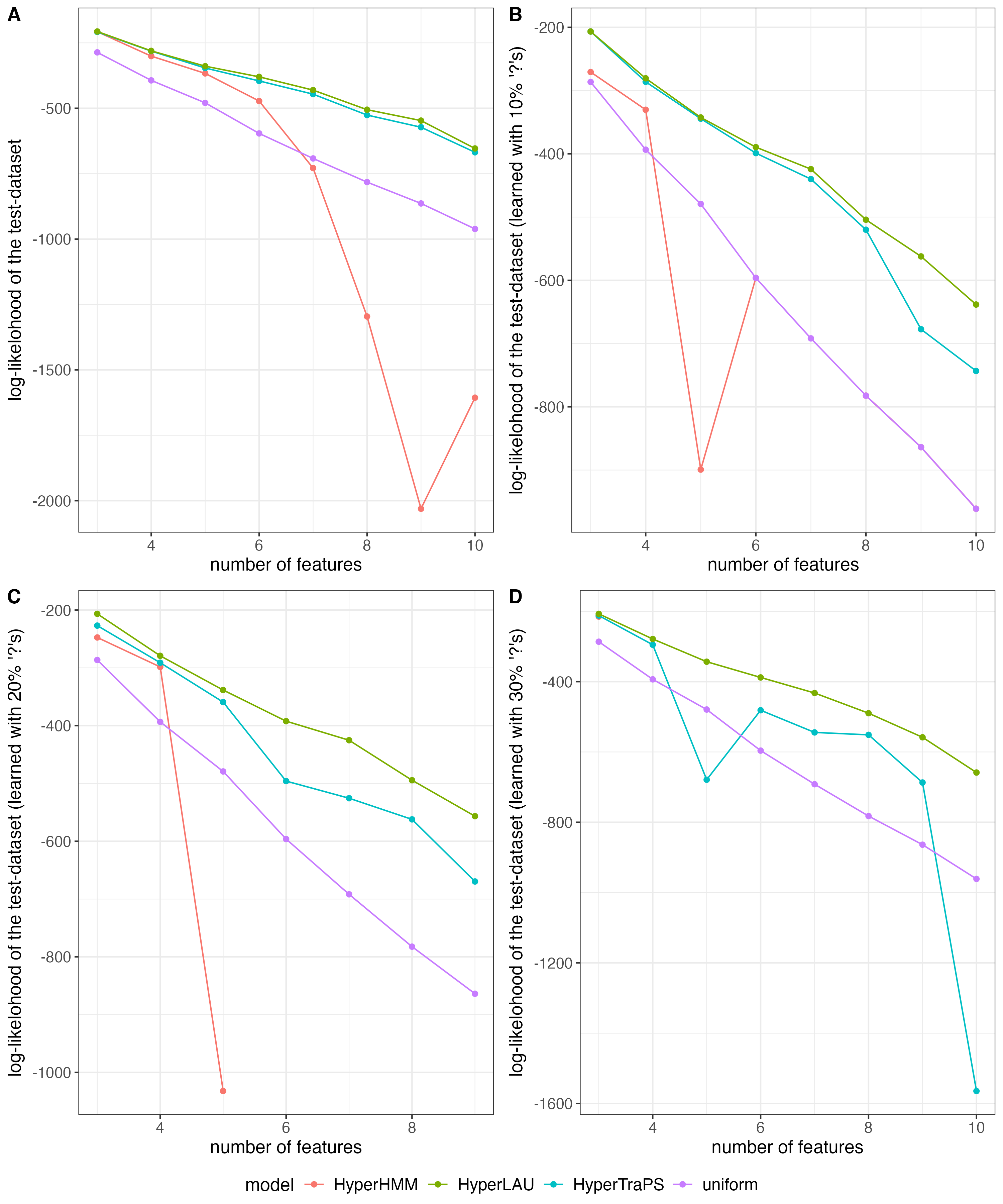}
	\caption{The development of the log-likelhood of the test-dataset under the outputs of the different models learned on the dataset (A) without uncertainties (B) 10\% uncertainties (C) 20\% uncertainties (D) 30\% uncertainties}
	\label{loglik_testset}
\end{figure}

When comparing the log-likelihood, as it is defined in this paper, of the data in the training set after inference of the different models, we can see that HyperHMM and HyperTraPS seem to provide a better fit to the data in this specific set. However, if one considers the log-likelihood of the separate test set under the learned parameters, a completely different picture emerges. In this case, HyperLAU performs as well as HyperTraPS, significantly better than under the uniform case, when every transition from a state is equally likely. HyperHMM seems to provide the most radical fit to the training data, so that data pairs which appear in the test set but not in the training set seem to be inferred with a very small probability, leading to a decrease in the log-likelhood drastically (see Figure \ref{loglik_testset}).  The difference in the log-likelihood of the test-dataset becomes even more pronounced when using the inferred parameters based on the datasets containing uncertainties (or the corresponding reduced ones). Here, we can see that HyperLAU performs also significantly better than HyperTraPS and provides the best log-likelihood values. 

\begin{figure}
	\centering
	\includegraphics[scale = 0.6]{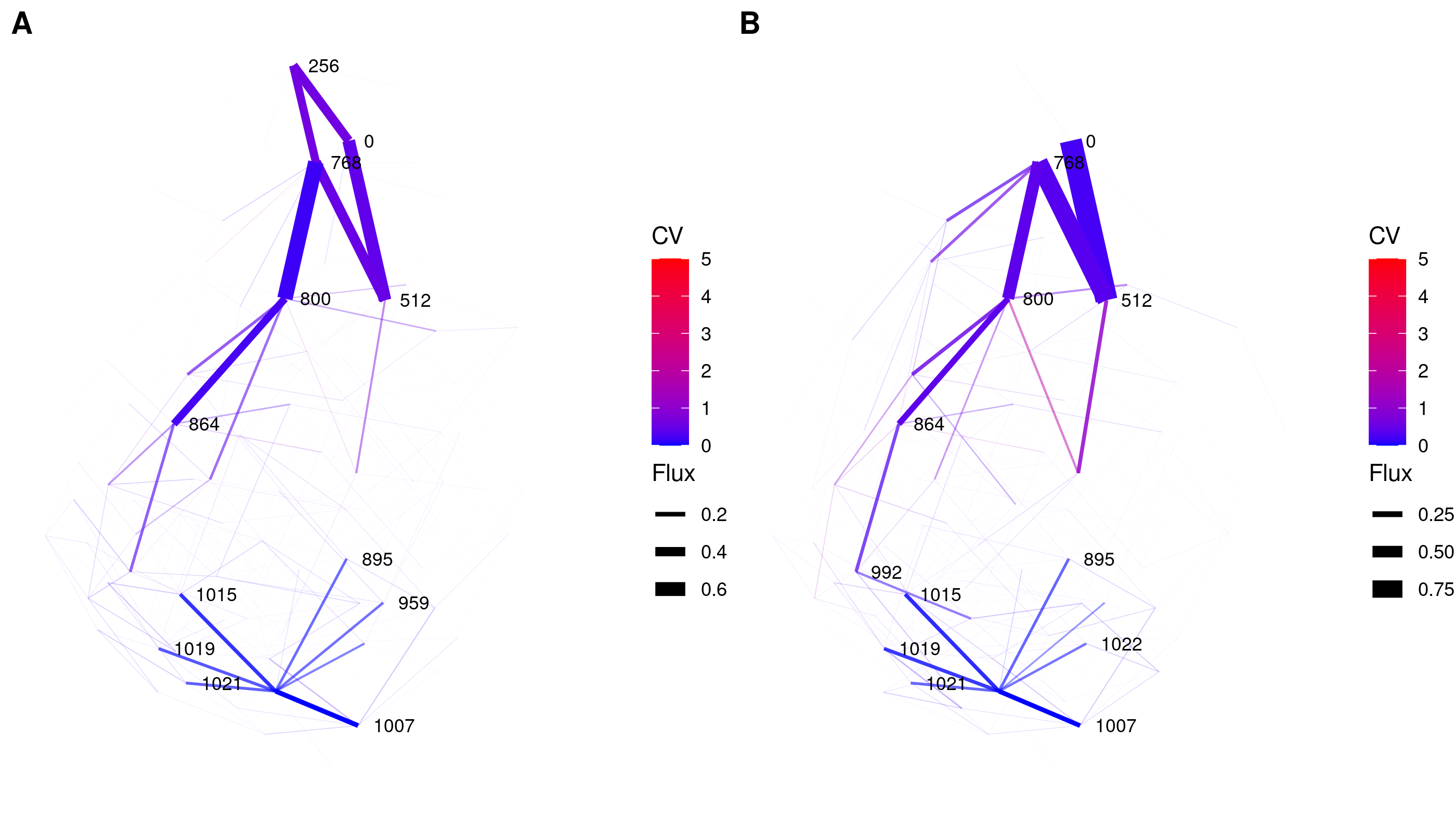}
	\caption{\small \textbf{Visualisation of evolutionary pathways learned by HyperLAU based on the tuberculosis dataset} \cite{casali_evolution_2014}. Plots show inferred transition networks through the evolutionary state space from 000... (top) to 111... (bottom) (as in Figure \ref{graph_abstr}D). In all plots, the thickness of the edges represents the probability flux between the corresponding state nodes (all coloured edges have minimum 0.05). Coefficient of variation (CV) is illustrated by the colour. Key states are labelled by the decimal representation of their binary labels (see below). \textbf{A}: Inference using original dataset with no uncertainties. \textbf{B}: Inference using artificially obscured dataset, where every position of the original dataset was replaced by a `?' with probability 0.5. Some key nodes: 256 = 0100000000 (RIF), 512 = 10000000000 (INH), 768 = 1100000000 (INH+RIF), 800 = 1100100000 (INH+RIF+STR), 864 = 1101100000 (INH+RIF+EMB+STR), 992 = 1111100000 (INH+RIF+PZA+EMB+STR). } \label{tb}
\end{figure}

\begin{figure}[h]
	\centering
	\includegraphics[scale = 0.9]{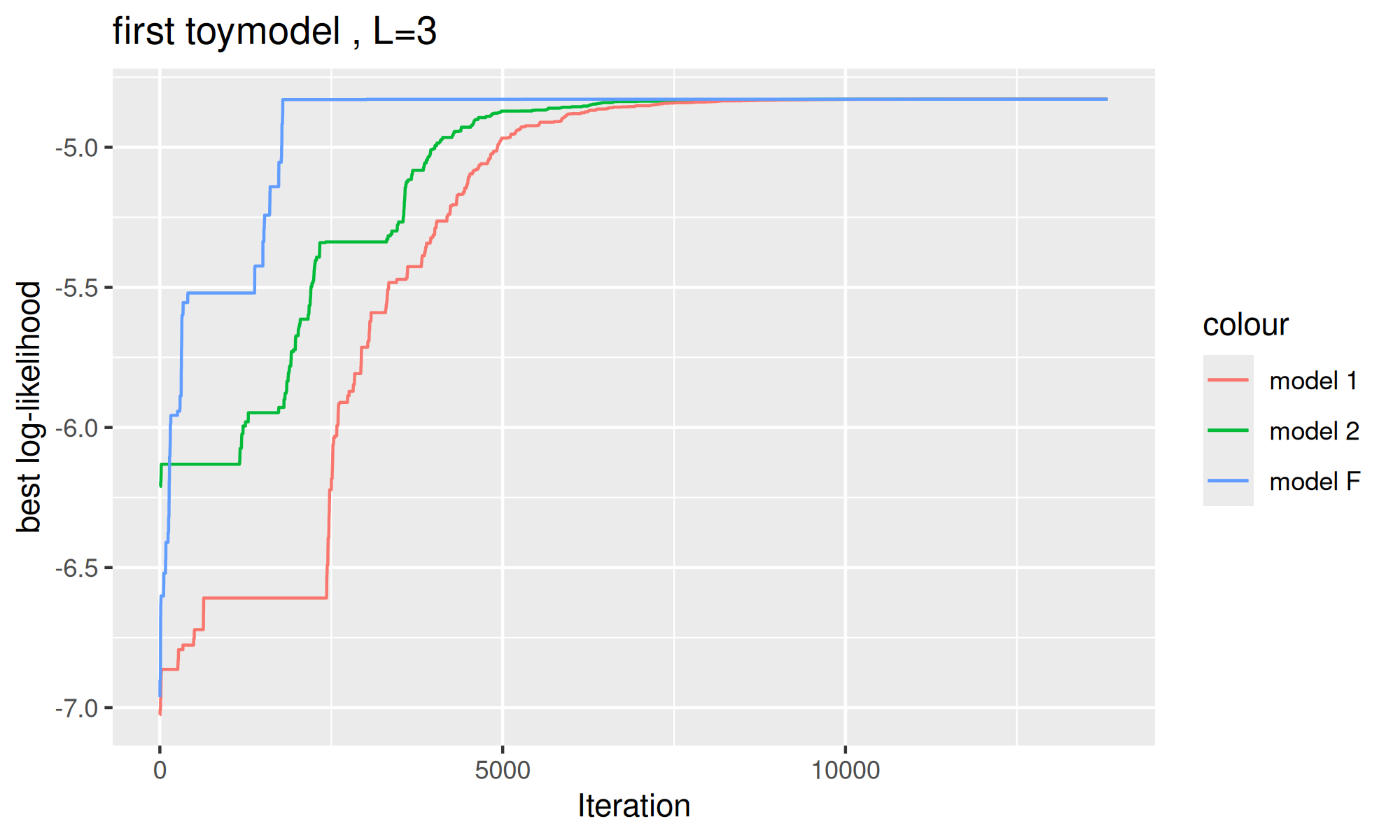}
	\caption{\small \textbf{Progression of the best log-likelihood of the first toy example during the SA process.} The iteration number on the x-axes indicates the iteration number in the optimization under model $F$.}
	\label{lik_toy1}
\end{figure}

\begin{figure}[h]
	\centering
	\includegraphics[scale = 0.9]{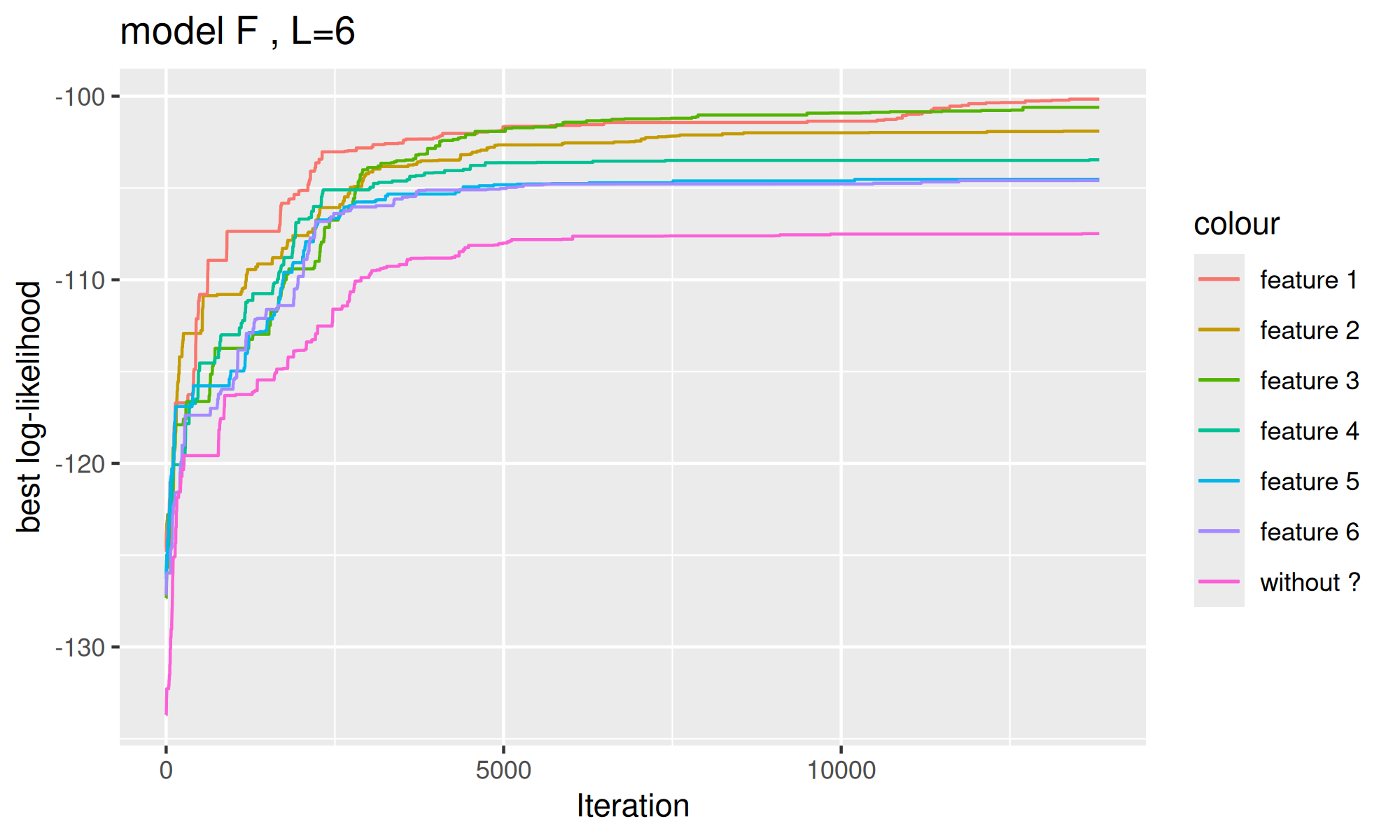}
	\includegraphics[scale=0.9]{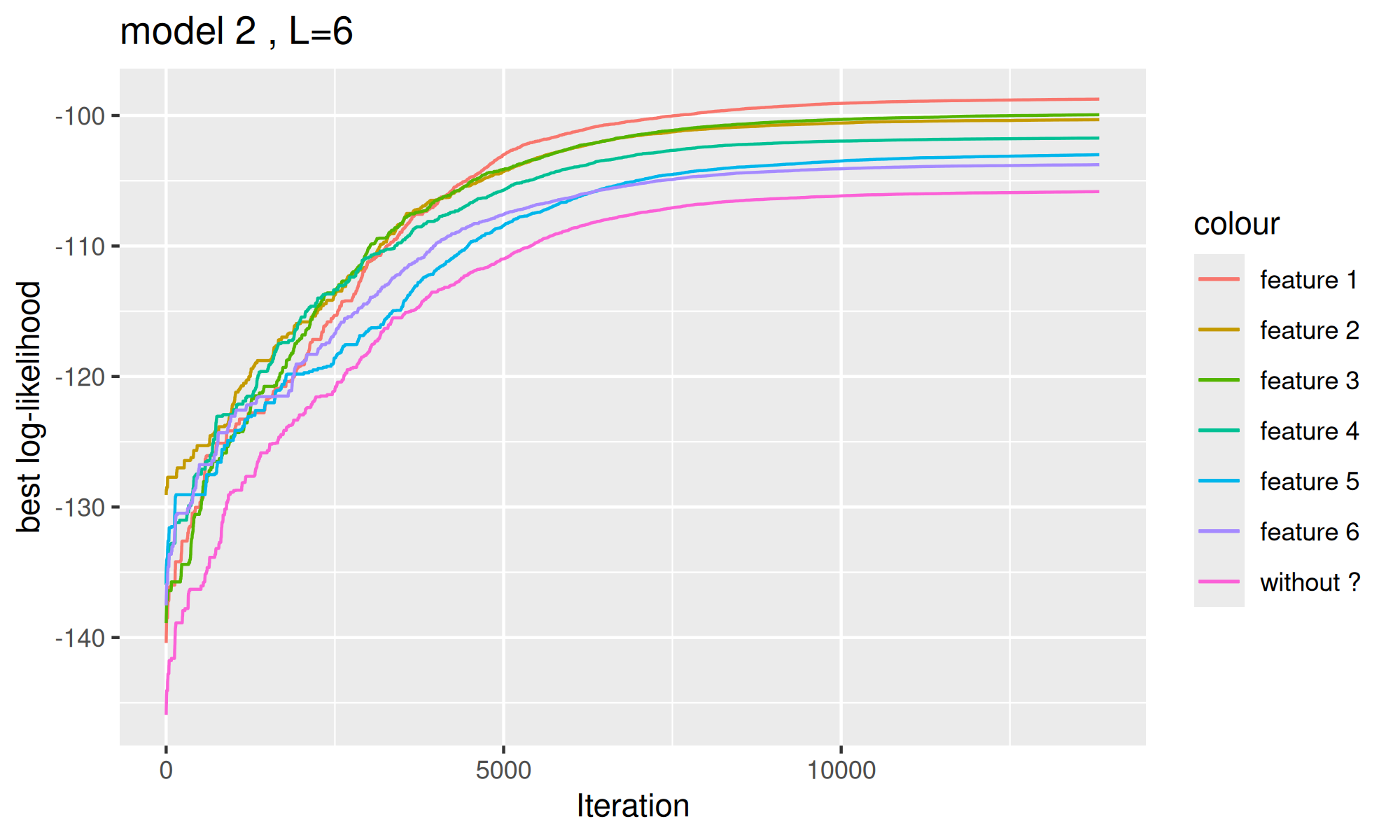}
	\caption{\small \textbf{Progression of the best log-likelihood of the second toy example during the SA process.} The iteration number on the x-axes indicates the iteration number in the optimization. The different colours of the trajectories indicate in which feature the uncertainty markers were inserted in the corresponding simulation. (top) Model $F$; (bottom) model 2.}
	\label{lik_features_model2}
\end{figure}

\begin{figure}
	\centering
	\includegraphics[scale = 0.2]{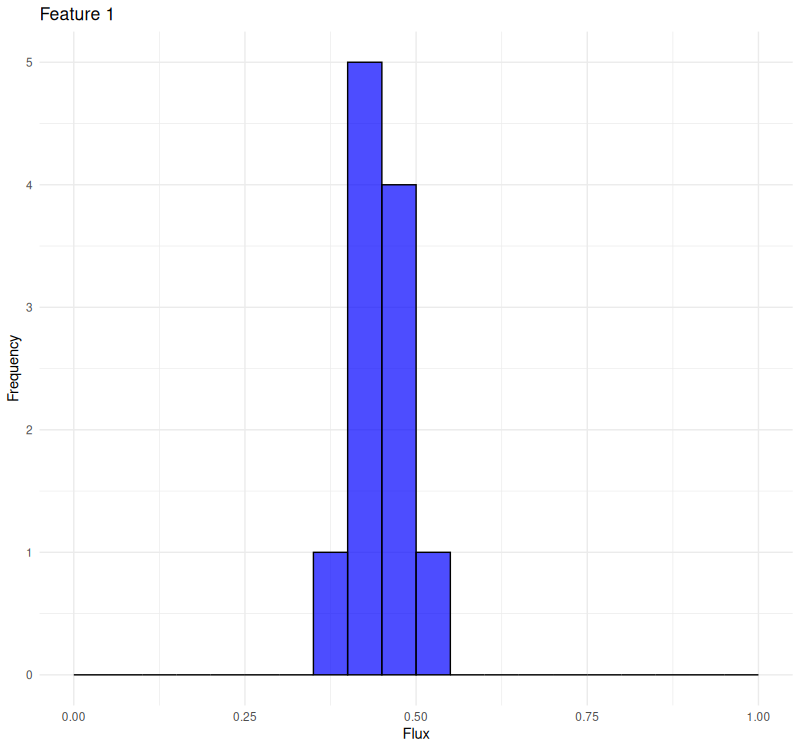}
	\includegraphics[scale = 0.2]{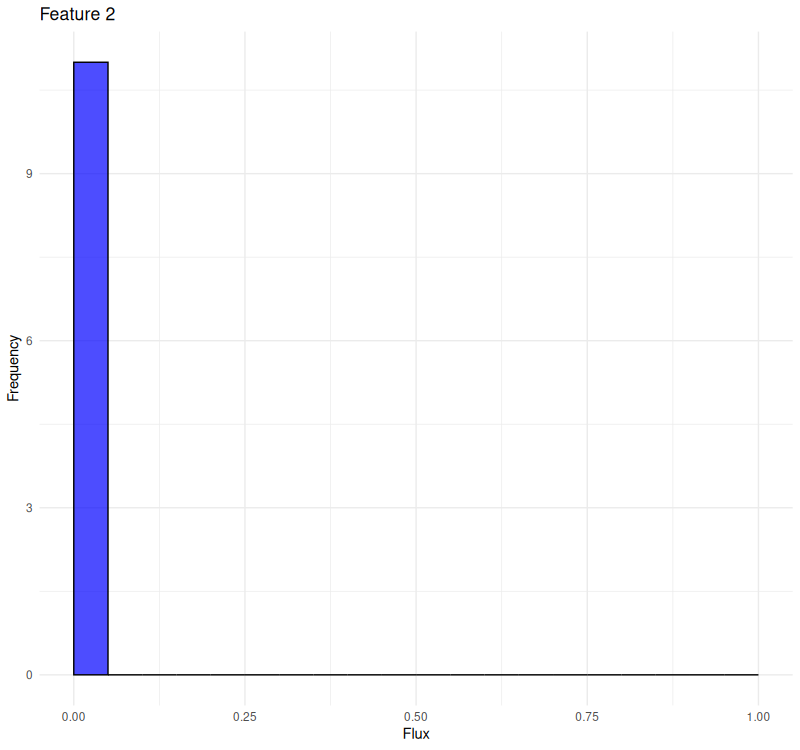}
	\includegraphics[scale = 0.2]{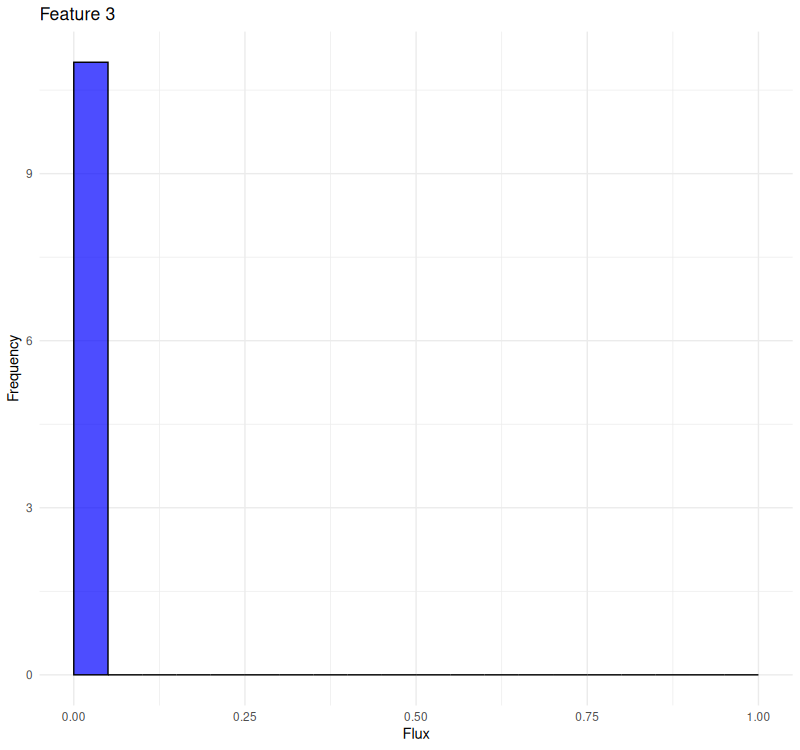}\\
	\includegraphics[scale = 0.2]{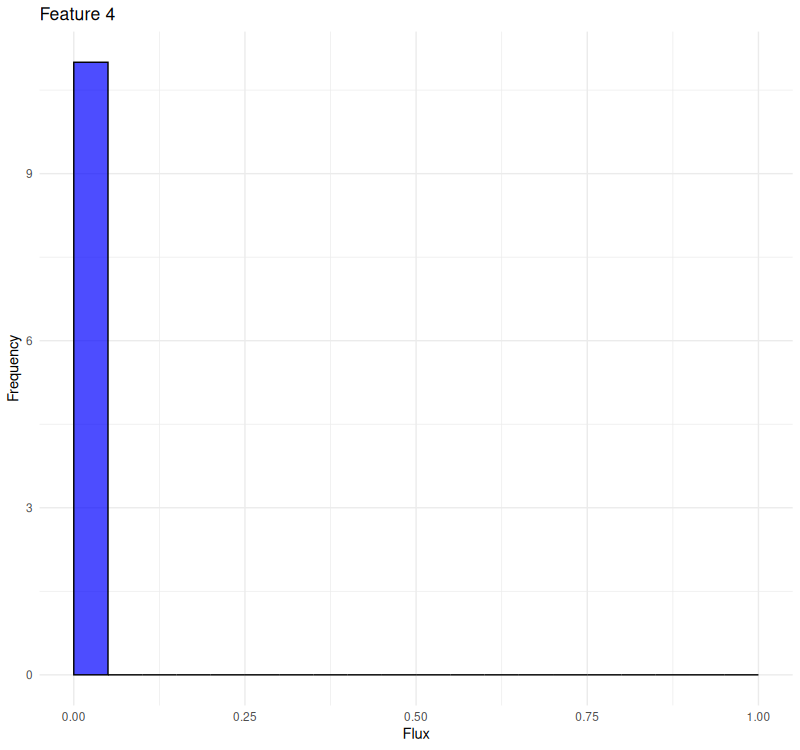}
	\includegraphics[scale = 0.2]{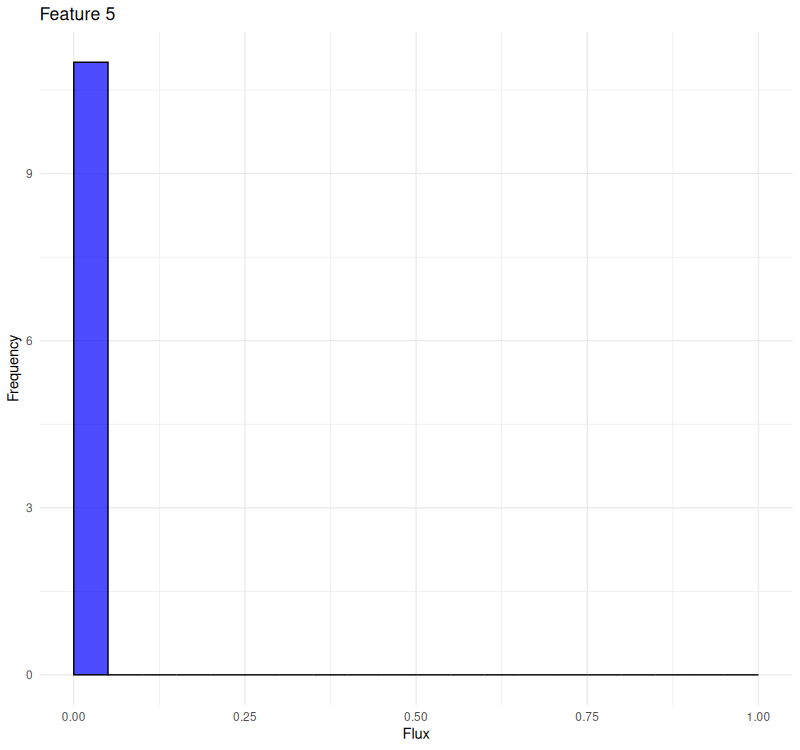}
	\includegraphics[scale = 0.2]{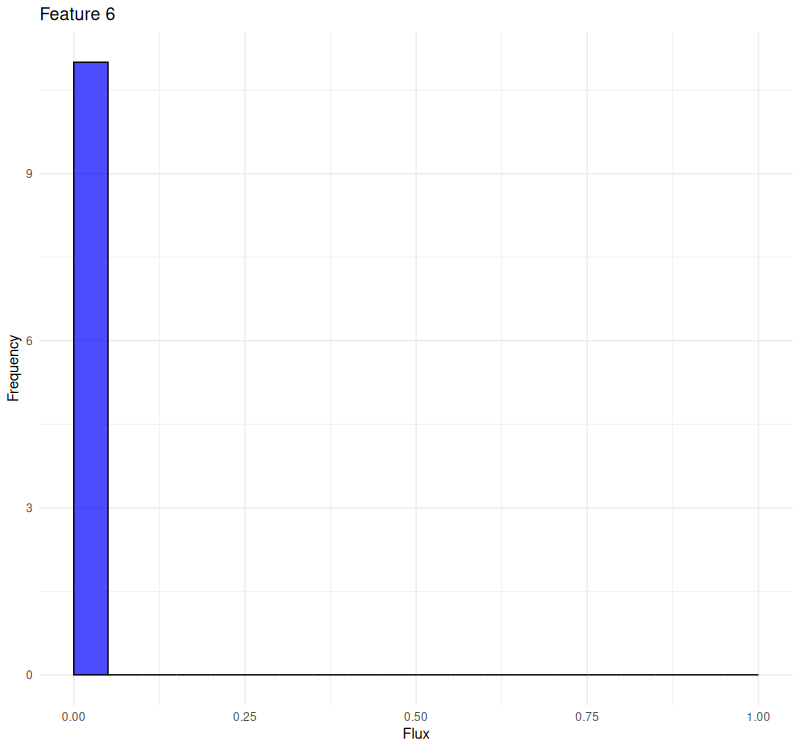}\\
	\includegraphics[scale = 0.2]{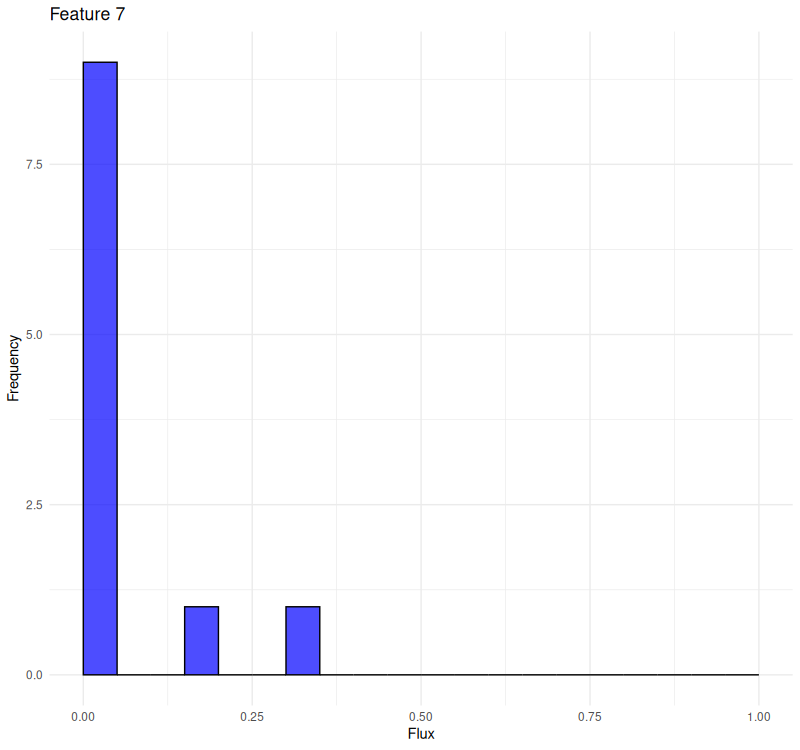}
	\includegraphics[scale = 0.2]{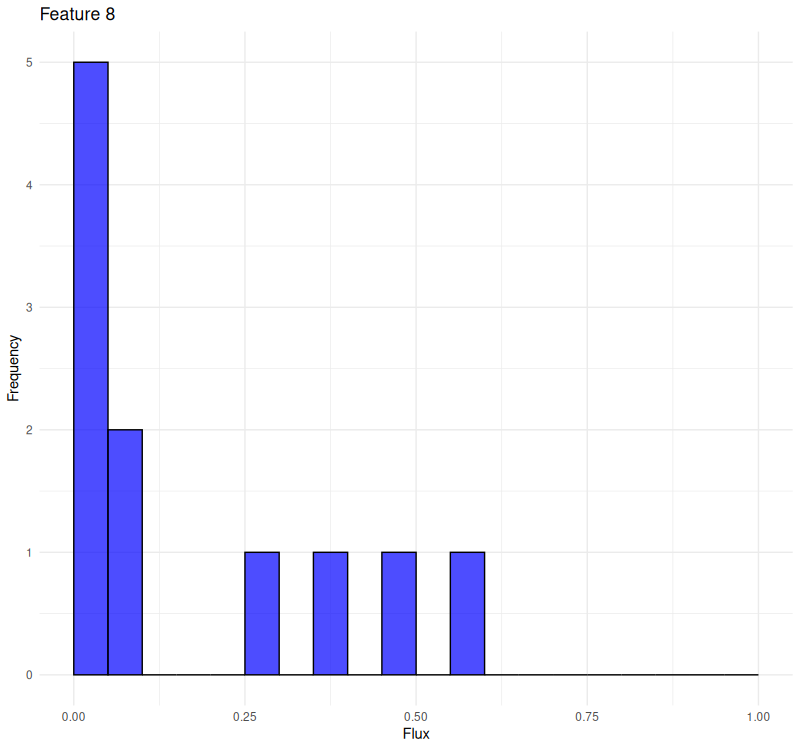}
	\includegraphics[scale = 0.2]{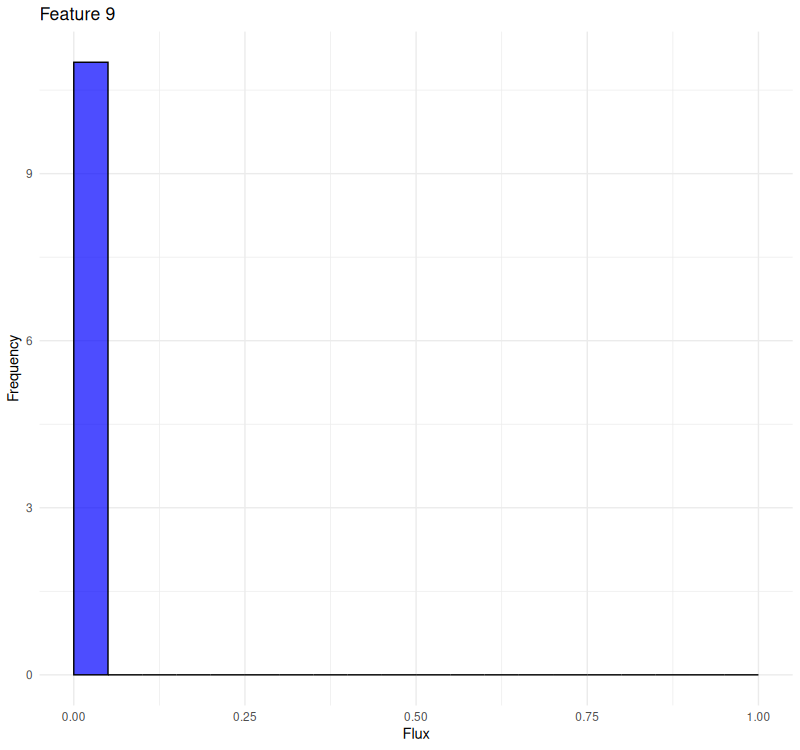}\\
	\includegraphics[scale = 0.2]{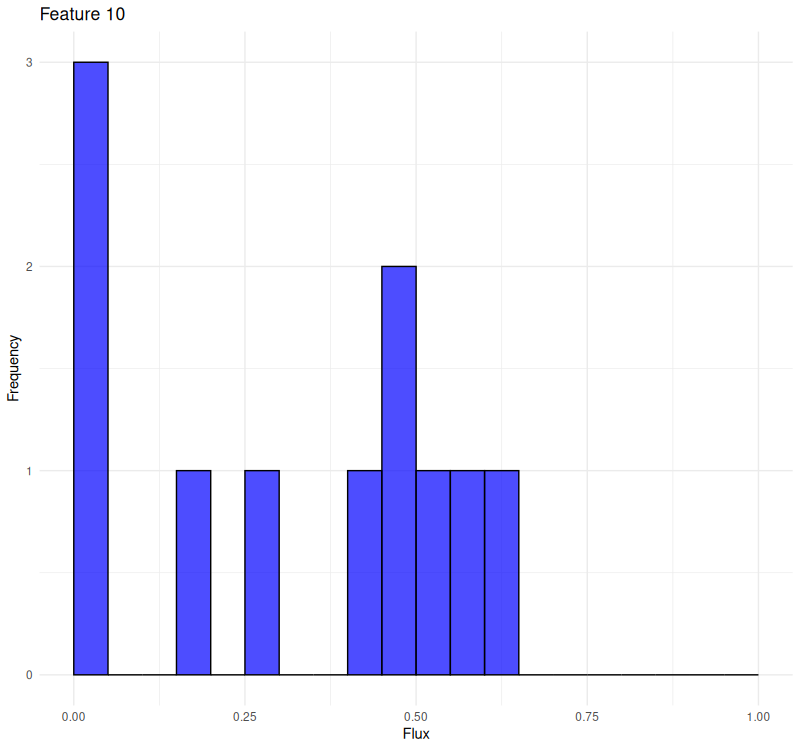}
	\caption{\small \textbf{Histograms of the different results during the bootstrapping.} Every histogram shows the probability distribution of one feature to be obtained in the first transition step.}
	\label{tb_features}
\end{figure}
\end{document}